\documentclass[12pt]{article}

\usepackage{amsmath,amssymb}
\usepackage{graphicx}
\usepackage{caption}
\usepackage{color}
\usepackage{dcolumn}
\usepackage{bm}
\usepackage[numbers,super,comma,sort&compress]{natbib}

\definecolor{background-color}{gray}{0.98}

\title{A general method to describe intersystem crossing dynamics in trajectory surface hopping
\thanks{This is the peer-reviewed version of the following article: \textit{S.\ Mai, P.\ Marquetand, L.\ Gonz\'alez: Int. J. Quant. Chem. 115, 1215--1231 (2015)}, which has been published in final form at http://dx.doi.org/10.1002/qua.24891. This article may be used for non-commercial purposes in accordance with Wiley Terms and Conditions for Self-Archiving.}
}

\author{Sebastian Mai\thanks{Institute of Theoretical Chemistry, University of Vienna, W\"ahringer Str. 17, 1090 Vienna, Austria}
\and Philipp Marquetand\thanks{corresponding author, email: philipp.marquetand@univie.ac.at} \footnotemark[1]
\and Leticia Gonz\'alez\thanks{corresponding author, email: leticia.gonzalez@univie.ac.at} \footnotemark[1]
}

\begin{document}

\maketitle


\begin{abstract}
Intersystem crossing is a radiationless process that can take place in a molecule irradiated by UV-Vis light, thereby playing an important role in many environmental, biological and technological processes. This paper reviews different methods to describe intersystem crossing dynamics, paying attention to semiclassical
trajectory theories, which are especially interesting because they can be applied to large systems with many degrees of freedom. In particular, a general trajectory surface hopping methodology recently developed by the authors, which is able to include non-adiabatic and spin-orbit couplings in excited-state dynamics simulations, is explained in detail. This method, termed \textsc{Sharc}, can in principle include any arbitrary coupling, what makes it generally applicable to photophysical and photochemical problems, also those including explicit laser fields. A step-by-step derivation of the main equations of motion employed in surface hopping based on the fewest-switches method of Tully, adapted for the inclusion of spin-orbit interactions, is provided. Special emphasis is put on describing the different possible choices of the electronic bases in which spin-orbit can be included in surface hopping, highlighting the advantages and inconsistencies of the different approaches.
\end{abstract}

\clearpage

\makeatletter
\renewcommand\@biblabel[1]{#1.}
\makeatother

\bibliographystyle{apsrev}

\renewcommand{\baselinestretch}{1.5}
\normalsize


\clearpage

\section*{\sffamily \Large Introduction}

\begin{figure}
  \centering
  \includegraphics[scale=1.0]{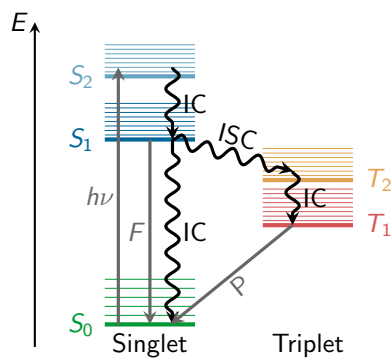}
  \caption{Jab{\l}onski diagram showing the conceptual photophysical processes. Straight arrows show radiative processes: absorption ($h\nu$), fluorescence (F), and phosphorescence (P); wavy arrows show radiationless processes: internal conversion (IC) and intersystem crossing (ISC). }
  \label{fig:jablonski}
\end{figure}

After a molecule is irradiated with light from the visible or UV parts of the spectrum, several photophysical processes can take place.\cite{Klessinger1995,McQuarrie1997,Turro2009}
These can be classified into radiative and radiationless processes and are best represented in a Jab{\l}onski diagram (see Figure~\ref{fig:jablonski}).
In radiative processes, the molecule interacts with light and absorbs (absorption) or emits (stimulated and spontaneous emission) a photon.
Emission can be further subdivided into fluorescence and phosphorescence, depending on whether the system emits from an electronic singlet or triplet state, respectively.
Radiationless processes are usually divided into internal conversion (IC), which is the transfer of population between electronic states of the same spin multiplicity, and intersystem crossing (ISC), which involves a change of the electronic spin. 
Oftentimes, these processes occur in cascades after the incidental absorption of a photon.
One typical chain of processes is absorption followed by IC to the lowest singlet state (as in Kasha's rule\cite{Kasha1950DFS}), followed by ISC to the triplet manifold, followed by IC to the lowest triplet state, from where the system may return to the electronic ground state by phosphorescence.

The non-radiative processes of IC and ISC are of fundamental importance in photochemistry and photobiology.
IC is ubiquitous in the non-adiabatic excited-state dynamics of polyatomic molecules because polyatomic molecules possess many degrees of freedom and a large number of electronic excited states, allowing for regions of degeneracy or close degeneracy in their potential energy surfaces (PES).
Intersections of two or more PES are usually refered to as conical intersections;\cite{Domcke2004} these conical intersections are precisely the funnels which facilitate efficient non-adiabatic transitions from an upper to a lower PES, which is IC.
IC is, for example, encountered in the process of visual perception, in particular during phototransduction.
During this process, the retinal chromophore of the opsin protein absorbs a photon and subsequently undergoes ultrafast IC facilitated by conical intersections, leading to photoisomerization\cite{Schoenlein1991S,Frutos2007PNAS} which changes the conformation of the protein and induces a neural signal.
Ultrafast IC is also responsible for avoiding harmful photochemical reactions of DNA\cite{Crespo-Hernandez2004CR,Barbatti2014TCC} and proteins,\cite{Weinkauf2002EPJD,Stephansen2012JACS,Perot2010JPCA,Vries2007ARPC}, protecting organisms from UV radiation.
IC is also present in the photochemistry of small molecules, for example O$_2$ and O$_3$,\cite{Seinfeld1997} SO$_2$,\cite{Heicklen1980RCI} NO$_2$\cite{Wilkinson2010ARPCSC} and other nitrogen oxides.\cite{Davis1993JPC}
The absorption of sunlight by these molecules and the subsequent photochemistry are central to atmospheric chemistry and the study of the environment and pollution.\cite{Seinfeld1997}

ISC is a key ingredient in molecular photophysics\cite{Klessinger1995,Turro2009} and hence it plays an important role in many areas of research.
For example, concepts like optical switching,\cite{Gutlich1994ACIEE} reverse saturable absorption\cite{Roy2000APL} or light-induced magnetism\cite{Epstein2003MB} are related to ISC in the field of molecular electronics.
The interconversion between electric energy and light (in artificial photosynthesis\cite{Ito2014ACIE,Segura2005CSR} and OLEDs\cite{Uoyama2012N}), magnetic data storage\cite{Epstein2003MB} and a number of biomimetic technologies\cite{Hsu2012JACS,Gust1993JACS} also rely on ISC.
ISC has also biochemical relevance as it is a key step in  many biochemical reactions (e.g., oxygen binding to carrier proteins~\cite{Saito2014JPCB}), harmful photochemical reactions of DNA strands, and has found application in phototherapies (e.g., photodynamical therapy\cite{Dolmans2003NRC}).
Furthermore, ISC is relevant to the study of combustion processes, where open-shell atoms (like oxygen) and radicals play a central role, in organic photochemistry, e.g., in $[2+2]$-cycloadditions,\cite{Griesbeck2004ACR} and in biomimetic catalysis.\cite{Zapata-Rivera2014CEJ}

Due to its increasing relevance in different disciplines, the number of studies on ISC has proliferated in the last years.
The study of photo-induced processes in transition-metal complexes,\cite{Roundhill1994} where ISC should be prominent due to the presence of the heavy metal atoms, has received special attention.
Chergui\cite{Chergui2012DT} has recently reviewed the photophysics of different transition metal complexes, paying particular attention to the timescale of ISC.
Intriguingly, the reported ISC timescales range from extremely fast, e.g., 30~fs for the $[\text{Fe}(\text{bpy})_3]^{2+}$ complex,\cite{Gawelda2007JACS} to about 800~ps in $\text{Pd}(\text{thpy})_2$.\cite{Yersin2004TCC}
This large spread is difficult to explain simply by the type of central atom, since for example the heavier Pd is expected to induce more efficient ISC than Fe.
Femtosecond time scales for ISC have been also observed in transition metal complexes based on ruthenium,\cite{Cannizzo2006ACIE} rhenium,\cite{Cannizzo2008JACSa} osmium,\cite{Bram2013JPCC} and other metals.\cite{Chergui2012DT}
More recently, also the role of ISC in organic molecules has captured the attention of researchers.
Examples of molecules where ISC is important range from aldehydes\cite{Amaral2010JCP} and small aromatic compounds like benzene,\cite{Parker2009CPL} naphthalene, anthracene and their carbonylic derivatives\cite{Cavaleri1996CPL,Ohshima2003JPCA,Satzger2004JPCA,Ramesdonk2006JPCA,Reichardt2009JCP,Yang2012JCP,Tamai1992CPL,Wolf2012M,Ma2007CEJ,Park2006JCP,Ou2013JPCC,Schalk2014JPCA,Huix-Rotllant2013PCCP} to nitrocompounds.\cite{Reichardt2009JCP,Vogt2013JPCA,Morales-Cueto2007JPCA,Zugazagoitia2008JPCA,Zugazagoitia2009JPCA,Lopez-Arteaga2013JPCB,Plaza-Medina2011JPCA,Mohammed2008JPCA,Collodo-Fregoso2009JPCA,Ghosh2012JPCA,Crespo-Hernandez2008JPCA}
Furthermore, ISC has been reported for thio-substituted,\cite{Reichardt2011JPCB,Reichardt2010JPCL,Reichardt2010CC,Pollum2014JCP,Martinez-Fernandez2012CC,Cui2014JPCL,Gobbo2014CTC} aza-substituted,\cite{Kobayashi2009JPCA} bromo-substituted\cite{Dietz1987JACS} and canonical nucleobases.\cite{Cadet1990,Schreier2007S,Barbatti2014CPC}

From a theoretical point of view, ISC can be explained by the interaction of states of different multiplicity by spin-orbit coupling (SOC), which is a relativistic effect.\cite{Marian2012WCMS}
It arises from the interaction of the magnetic field induced by the orbital momentum of an electron and the magnetic dipole related to the intrinsic electron spin.
Since the orbital momentum depends on the nuclear charges, SOC is stronger in molecules possessing heavy atoms.
For systems lacking heavy atoms, SOC has been historically regarded as negligible and hence ISC was believed to show much smaller reaction rates than IC.\cite{McQuarrie1997}
However, the actual ISC rate depends not only on the size of the SOCs, but also on the density of states, and the particular shape of the excited-state potential energy landscapes, most importantly the location and accessibility of surface crossing regions (e.g., between singlets and triplets).
This is the reason why an \emph{a priori} assessment of ISC rates is difficult and experiments have found that ISC can occur on very different time scales (from fs to ns), apparently not correlated with the magnitude of SOCs.
In this context, theoretical methodologies are particularly useful to disclose the factors that determine ISC in a general framework.

For relatively slow ISC processes, the calculation of ISC rate constants using Landau-Zener theory,\cite{Grimme1998CPL} Fermi's golden rule\cite{Harvey2000CPL,Tatchen2007PCCP} or Marcus-Jortner theory\cite{Harvey2000CPL,Cui1999JCP,Veldman2008JPCA} is well established.
For ultrafast ISC processes -- where the system is very far from equilibrium -- a dynamical treatment describing the motion on the PES and taking into account the kinetic processes that lead the system towards the relevant crossing region is mandatory.
Encompassing the experimental discoveries involving ISC, the last decades have seen a surge of excited-state nuclear dynamics studies describing ISC.
A number of studies have used quantum wavepacket dynamics, where inclusion of SOC is straightforward; however, these studies are usually restricted to few degrees of freedom.
Examples include dihalogens in argon matrices,\cite{Bargheer2002PCCP,Korolkov2004JCP,Korolkov2004CPL} transition-metal complexes,\cite{Daniel1995JCP,Heitz1995CCR,Daniel1996IJQC,Daniel1999JPCA,Bruand-Cote2002CEJ,Amor2007CP,Costa2008NJC,Heydova2012JPCA,Gourlaouen2014DT,Brahim2014CTC,Kayanuma2011CCR,Ando2012CPL,Capano2014JPCA} and collision reactions.\cite{Zhao2013JCP}
Further quantum dynamical studies investigated ISC in benzene,\cite{Minns2010PCCP} hydrogen fluoride,\cite{Cohen2007CPL} and sulphur dioxide.\cite{Leveque2014JCP_ISC}
Full-dimensional dynamical studies can be performed using extensions of the trajectory surface hopping methodology.
Different schemes have been employed to investigate ISC in collision reactions like $\text{O}+\text{C}_2\text{H}_4$,\cite{Fu2012JCP,Fu2012PNAS,Rajak2014JCP,Hu2008JPCA} $\text{O}+\text{N}_2$,\cite{Tachikawa1997JPCA} $\text{Na}+\text{HCl}$,\cite{Yamashita1990CPL} or $\text{S}+\text{H}_2$\cite{Maiti2004JPCA} as well as in the dynamics of O$_2$ on Al surfaces.\cite{Carbogno2010PRB}
Trajectory surface hopping was also used to treat ISC in molecules like sulphur dioxide,\cite{Mai2014JCP_SO2} acrolein,\cite{Cui2014JCP} acetone,\cite{Favero2013PCCP} pentanal,\cite{Shemesh2013JPCA} 2-butene,\cite{Warshel1975CPL} 6-thioguanine,\cite{Martinez-Fernandez2014CS} cytosine,\cite{Richter2012JPCL,Mai2013C} uracil,\cite{Richter2014PCCP} as well as few transition-metal complexes.\cite{Tavernelli2011CP,Freitag2014IC}

The present contribution aims at describing and reviewing the available approaches described in the literature to treat ISC in trajectory surface hopping methods, paying special attention to the general surface hopping method recently developed by the authors,\cite{Richter2011JCTC,Mai2014SHARC} which allows treating non-adiabatic IC and ISC on the same footing.
This methodology is especially well suited to investigate ultrafast ISC (on timescales up to a few ps), because long timescales are usually too expensive for nuclear dynamics approaches.
In this regard, our methodology can be considered complementary to other non-dynamical treatments of ISC,\cite{Harvey2000CPL,Tatchen2007PCCP,Cui1999JCP,Veldman2008JPCA} which are not well applicable to ultrafast processes, but work well for slower reactions.
Our methodology has been recently implemented in a completely new \textsc{Sharc} surface hopping dynamics code.
The \textsc{Sharc} program suite, which contains the dynamics code itself and a number of setup, interface and analysis tools, is available online free of charge.\cite{Mai2014SHARC}

The rest of the paper is then organized as follows.
The next section introduces generally the surface hopping method and describes different ways in which SOC has been introduced to account for ISC.
Then, section~3 describes step-by-step a way how to arrive at a generally applicable, computationally feasible and satisfactorily accurate methodology for the description of excited-state molecular dynamics including ISC.
Finally, section~4 summarizes the presented methodology and gives an outlook on future developments in this field.


\section*{\sffamily \Large Dynamics Simulations of Intersystem Crossing}

The basic equation for excited-state dynamics simulations is the time-dependent Schr\"odinger equation, where the Hamiltonian includes all non-adiabatic and potential couplings between the electronic states of interest.
Simulating the motion of a quantum wavepacket according to this Hamiltonian by numerically integrating the time-dependent Schr\"odinger equation is usually termed quantum dynamics.\cite{Meyer2009,Tannor2006}
This approach includes all quantum-mechanical effects like tunneling, coherence and the splitting of the wavepacket and can deliver quantitative results given accurate PES.\cite{Meyer2009}
However, the necessity to know \emph{a priori} the full multi-dimensional PES is a severe bottleneck of the method, rendering quantum dynamics in full dimensionality unfeasible for systems with many degrees of freedom.
There exist a number of approximations that allow treating an increasing number of degrees of freedom, e.g.\ the ``multi-configurational time-dependent Hartree'' method (MCDTH),\cite{Meyer2009} its variants ``Gaussian-based MCTDH'',\cite{Burghardt1999JCP} ``multilayer MCTDH''\cite{Wang2003JCP} and ``variational multi-configurational Gaussians'',\cite{Lasorne2006CPL} and the unrelated ``full multiple spawning'' method.\cite{Martinez1996JPC} 
Alternatively, the dimensionality problem can be tackled with trajectory surface hopping non-adiabatic dynamics.

The concept of surface hopping was originally devised by Tully\cite{Tully1971JCP} and greatly improved later by his ``fewest-switches criterion''.\cite{Tully1990JCP}
Given its wide acceptance and broad applicability, it has been reviewed extensively since then, see for example references\cite{Doltsinis2002JTCC,Barbatti2011WCMS,Persico2014TCA,Malhado2014FC}.
Surface-hopping approximates the motion of the nuclear wavepacket on the excited-state PES by an ensemble of a large number of independent, classical trajectories.
The nuclear motion is determined by Newton's equations, where the force acting on the nuclei is the gradient of one particular electronic state (the \emph{active} state).
Non-adiabatic population transfer between the electronic states can lead to instantaneous switches of the active state.
This switching -- the eponymous ``hopping'' -- is conducted stochastically based on the rate of change of the electronic populations\cite{Tully1990JCP} and is accompanied by an adjustment of the kinetic energy in order to preserve the total energy.

Surface-hopping methods can deliver a wealth of information about the excited-state dynamics of molecules.
The ensemble ansatz allows obtaining branching ratios between different reaction channels.
Time-dependent excited-state populations indicate the evolution of the system across different states, enabling to propose kinetic models.
The geometries at which surface hops occurred can be used to infer surface crossing points, both conical intersections and singlet-triplet crossings (or crossing points between states of any multiplicity).
By calculating observables along with the trajectories and taking the ensemble average, it is even possible to directly correlate the results with experimental observations.

Compared to other dynamics methodologies, surface hopping has a number of advantages,\cite{Barbatti2011WCMS} which justify its popularity:
(i) The classical mechanics ansatz for the nuclear motion makes the method conceptually simple and easy to implement.
(ii) The trajectories need only local information on the electronic states (i.e., energies, gradients, couplings).
Thus, surface hopping allows for on-demand (``on-the-fly'') calculation of the electronic properties, which makes a treatment of all nuclear degrees of freedom possible without the need to construct the full multi-state PES beforehand.
(iii) Each trajectory is independent from the rest of the ensemble.
Hence, the trajectories can be trivially parallelized.

The fact that all degrees of freedom can be included in the simulation allows surface hopping to describe large systems, which is likely the reason of its widespread use.
Another strength of on-the-fly methods is that they allow going beyond the so-called linear vibronic coupling\cite{Koppel1978JCP, Koeppel2007} and linear vibronic spin-orbit coupling\cite{Henry1971JCP,Tatchen2007PCCP} approaches, where the relevant matrix elements are only approximated by some truncated power series around a reference geometry.
In trajectory surface hopping methods, the availability of the full geometry dependence of the electronic properties makes it automatically possible to describe effects like the promotion of SOC by excitation of symmetry-reducing vibrational modes.
Unfortunately, the classical approximation has the price that some quantum-mechanical effects are not treated properly.
For example, nuclear vibration is not quantized, which means that nuclear interference effects are neglected and zero-point energy is formally not treated correctly.
Additionally, because of the independent-trajectory ansatz, quantum coherence between the electronic states is not properly taken into account.
The latter effect can be partially mitigated using so-called decoherence corrections; the interested reader is referred to the according literature.\cite{Granucci2007JCP,Granucci2010JCP,Cantatore2014CTC,Bajo2014JCP,Zhu2004JCP,Zhu2005JCTC,Jasper2005JCP,Jasper2007JCP,Cheng2008JCP,Prezhdo1997JCPa,Jaeger2012JCP,Subotnik2011JCP,Shenvi2011JCP,Subotnik2013JCP,Jakubetz1999PCCP,Bedard-Hearn2005JCP,Nelson2013JCP,Yonehara2012CR,Malhado2014FC}
Tunneling effects can also not be described with surface hopping although there exist several approaches to alleviate this deficiency, see, e.g., references\cite{Makri1989JCP, Hammes-Schiffer1994JCP, Takatsuka1999PR, Xing2001JPCB, Larregaray2002PCCP}.

The original surface hopping method was formulated to consider only non-adiabatic couplings\cite{Tully1990JCP,Hammes-Schiffer1994JCP} between electronic states of the same multiplicity and has been therefore intensively used to describe excited state dynamics involving IC.
By incorporating SOC into the surface hopping procedure, dynamical studies can be extended to also describe ISC.
In the following, we list the different concepts which -- to our knowledge -- have been described in the literature, ordered by how consistently ISC is described with regard to IC:
\begin{itemize}
  \item On the simplest level, hopping between electronic states of different multiplicity is performed ``manually''.
  This can be done by exploring first the deactivation mechanism within, e.g., the singlet manifold with regular surface hopping and then identifying singlet-triplet crossing regions, where ISC is assumed to occur.
  These regions then serve as a starting point for an independent simulation in the triplet manifold.
  This approach was used in the 70s by Warshel and Karplus,\cite{Warshel1975CPL} but is still in use nowadays for the description of exceptionally difficult systems.\cite{Tavernelli2011CP,Shemesh2013JPCA,Freitag2014IC}
  \item On the next level, regular surface hopping can be combined with hopping probabilities between states of different multiplicities calculated only at crossing points based on Landau-Zener theory.\cite{Zener1932PRSA}
  Although in this way hoppings due to IC and hoppings due to ISC are treated inconsistently, this approach has been used to some extent in describing collision reactions.\cite{Rajak2014JCP,Tachikawa1997JPCA,Hu2008JPCA,Yamashita1990CPL}
  \item A further improvement is the proper inclusion of spin-orbit couplings in the equation of motion which governs the evolution of the electronic populations and hence the hopping probabilities.
  This allows in principle to have ISC in every time step of the simulation, not just at singlet-triplet crossings.
  In the simplest variant, the propagation is carried out using electronic states which are eigenstates of the total spin ($\hat{S}^2$) and spin projection ($\hat{S}_z$) operators.
  In this paper we term these states ``MCH'' states (referring to the ``molecular Coulomb Hamiltonian'', see below for a detailed discussion), while in the literature they are also referred to as ``spin-diabatic'' states.\cite{Granucci2012JCP}
  Surface-hopping using these states has been adopted in several cases, either assuming a constant value for the SOC matrix elements\cite{Fu2012JCP,Fu2012PNAS} or with geometry-dependent SOC.\cite{Carbogno2010PRB,Cui2014JCP,Curchod2013C,FrancodeCarvalho2014JCP}
  \item Finally, the most advanced schemes to describe ISC employ a diagonalization of the electronic Hamiltonian including SOC, and basically conduct Tully's surface hopping using the obtained eigenstates.
  Because the full electronic Hamiltonian is diagonal in the basis of these states, here we label it the ``diagonal'' basis (also called ``spin-adiabatic'' elsewhere\cite{Granucci2012JCP}).
  The diagonal electronic basis offers a number of advantages, which will be discussed in detail below.
  This approach has been used already in 2004 to describe the $\text{S}+\text{H}_2$ reaction\cite{Maiti2004JPCA} and recently saw a number of applications.\cite{Favero2013PCCP,Granucci2012JCP,Martinez-Fernandez2014CS,Richter2011JCTC,Richter2012JPCL,Mai2013C,Mai2014JCP_SO2,Richter2014PCCP} 
  A general on-the-fly implementation of surface hopping in this fashion is available in the \textsc{Sharc} suite.\cite{Richter2011JCTC,Mai2014SHARC}
\end{itemize}
In passing we note that similar methodologies using the equivalents of the MCH\cite{Gai1992JAP,Mitric2009PRA} or diagonal\cite{Thachuk1996JCP,Jones2008JPCA,Richter2011JCTC,Marquetand2011FD,Bajo2011JPCA} electronic bases have also been developed in different contexts to describe other processes, like interactions of electromagnetic fields with molecular dipole moments.

The following section is devoted to explain exclusively the two last approaches to ISC surface hopping, since only these offer a consistent description of ISC versus IC.


\section*{\sffamily \Large Surface Hopping Methodology for Intersystem Crossing}

In the following, we provide a step-by-step derivation of the main working equations necessary to perform trajectory surface hopping simulations focussing on the application to ISC problems.
We start giving an overview over the workflow of a surface hopping simulation.
Then, we develop the equation of motion for the electronic wavefunction, which will be our main concern here, since ISC arises from coupling of electronic states of different multiplicity.
Afterwards, we discuss which electronic basis functions should be used in the propagation of the wavefunction, and the implications of this choice, in particular with respect to the integration of the equations of motion and the gradients used for the nuclear dynamics.

\subsection*{Surface Hopping}

Surface-hopping is a special type of molecular dynamics (MD).
In MD, the nuclear motion -- which should be ideally described by quantum mechanics -- is approximated with classical mechanics.
The nuclear equation of motion is in this case given by Newton's second law, which for the nuclear position $\mathbf{R}_A$ of atom $A$ reads as:
\begin{equation}
  m_A\frac{\mathrm{d}^2}{\mathrm{d} t^2}\mathbf{R}_A(t)=-\nabla_A E_\beta(\mathbf{R}(t)).
  \label{eq:newton}
\end{equation}
Here, $m_A$ is the mass of atom $A$ and $-\nabla_A E_\beta(\mathbf{R}(t))$ is the gradient of the energy of the electronic state $\beta$ (we shall call this state the \emph{active} state) with respect to the position of atom $A$; the energy depends on the geometry $\mathbf{R}$ of all atoms.
In the standard formulation of MD, at each time step only the gradient $-\nabla_A E_\beta$ needs to be evaluated, e.g., by electronic structure methods, and performing the MD simulation consists of solely integrating equation~\eqref{eq:newton}.
Integration from time $t$ to time $t+\Delta t$ can be effectively done with the velocity-Verlet algorithm,\cite{Verlet1967PR} which can be expressed by the following equations:
\begin{align}
  \mathbf{a}_A(t)=&
    -\frac{1}{m_A}\nabla_AE_\beta(\mathbf{R}(t)),\label{eq:vv:a1}\\
  \mathbf{R}_A(t+\Delta t)=&
    \mathbf{R}_A(t)+\mathbf{v}_A(t)\Delta t + \frac{1}{2}\mathbf{a}_A(t)\Delta t^2,\label{eq:vv:r}\\
  \mathbf{a}_A(t+\Delta t)=&
    -\frac{1}{m_A}\nabla_AE_\beta(\mathbf{R}(t+\Delta t)),\label{eq:vv:a2}\\
  \mathbf{v}_A(t+\Delta t)=&
    \mathbf{v}_A(t)+\frac{1}{2}\left[\mathbf{a}_A(t)+\mathbf{a}_A(t+\Delta t)\right]\Delta t,\label{eq:vv:v}
\end{align}
where $\mathbf{a}_A$ is the acceleration of atom $A$ and $\mathbf{v}_A$ is the velocity of atom $A$.

In order to incorporate non-adiabatic effects by means of the surface hopping scheme, a prescription of how to choose the correct active state $\beta$ at each time step is required.
This choice is based on the evolution of an electronic wavefunction along the nuclear trajectory.
Accordingly, for  every time step the following steps need to be carried out:
\begin{enumerate}
  \item For the current nuclear position $\mathbf{R}(t)$, the necessary electronic properties (e.g., electronic Hamiltonian and non-adiabatic couplings) are determined.
  \item From these properties, the electronic wavefunction $\Psi(t)$ is propagated.
  \item From the electronic wavefunction, the hopping probabilities $\mathbf{h}$ are calculated and the active state $\beta$ is determined stochastically.
  \item The gradient of $E_\beta$ is used to calculate the new geometry $\mathbf{R}(t+\Delta t)$.
\end{enumerate}
Based on this scheme, it can be seen that the evolution of $\Psi$ and $\mathbf{R}$ are coupled.
In the following subsection, we shall focus on the evolution of $\Psi$ and the hopping probabilities $\mathbf{h}$ which can be derived from $\Psi$.

\subsection*{Electronic Equation-of-Motion}

We expand the electronic wavefunction $|\Psi\rangle$ (we omit the explicit time-dependence for brevity) by using $\sum\limits_\alpha |\psi_\alpha\rangle\langle\psi_\alpha|=1$:
\begin{equation}
  |\Psi\rangle
  = \sum\limits_\alpha |\psi_\alpha\rangle\langle\psi_\alpha|\Psi\rangle,
 \label{eq:wavefunction}
\end{equation}
where $\alpha$ runs over all electronic states in the \emph{model space}.
In order to arrive at a computationally tractable problem, one usually includes only a few states in the model space (e.g.\ the few lowest electronic states of each multiplicity for simulations of ISC dynamics), chosen in a way that the model space describes the total electronic wavefunction reasonably well for the processes under investigation.

In order to obtain the electronic equation of motion, we insert expression \eqref{eq:wavefunction} into the electronic time-dependent Schr\"odinger equation,
\begin{equation}
  \text{i}\frac{\mathrm{d}}{\mathrm{d} t}|\Psi\rangle = \hat{H}_\text{el}|\Psi\rangle
\end{equation}
and project on $\langle\psi_\beta|$:
\begin{equation}
  \text{i}
  \sum\limits_\alpha
  \langle\psi_\beta|\frac{\mathrm{d}}{\mathrm{d} t}|\psi_\alpha\rangle\langle\psi_\alpha|\Psi\rangle
  +
  \langle\psi_\beta|\psi_\alpha\rangle\frac{\mathrm{d}}{\mathrm{d} t}\langle\psi_\alpha|\Psi\rangle
  =
  \sum\limits_\alpha
  \langle\psi_\beta|\hat{H}_\text{el}|\psi_\alpha\rangle\langle\psi_\alpha|\Psi\rangle.
\end{equation}
Here, $\hat{H}_\text{el}$ is the electronic Hamiltonian operator, which we will define more precisely below.
Note that in the following we refer to the electronic Hamiltonian simply by the ``Hamiltonian'', as nuclei are treated classically and hence there is no nuclear Hamiltonian.

By using $\langle\psi_\beta|\psi_\alpha\rangle=\delta_{\beta\alpha}$, and after rearranging and dividing by $\text{i}$, one obtains:
\begin{equation}
  \frac{\mathrm{d}}{\mathrm{d} t}\langle\psi_\beta|\Psi\rangle
  =
  -\sum\limits_\alpha
  \left[
    \text{i}\langle\psi_\beta|\hat{H}_\text{el}|\psi_\alpha\rangle
    +
    \langle\psi_\beta|\frac{\mathrm{d}}{\mathrm{d} t}|\psi_\alpha\rangle
  \right]
  \langle\psi_\alpha|\Psi\rangle.
  \label{eq:eom_general}
\end{equation}
This equation is generally valid regardless the particular choice of the (complete) basis $|\psi_\alpha\rangle$, which we will call the ``representation'' in the following. 
The matrix representation of the Hamiltonian in a given basis ``rep'' will be denoted by bold $\mathbf{H}^\text{rep}$ in the remainder of the text, and the Hamiltonian matrix elements by $H_{\beta\alpha}^\text{rep}=\langle\psi_\beta^\text{rep}|\hat{H}_\text{el}|\psi_\alpha^\text{rep}\rangle$.
For simplicity in the notation, the subscript ``el'' will by omitted in all electronic Hamiltonian matrices and matrix elements.
We also define the wavefunction coefficient vector $\mathbf{c}^\text{rep}$ containing the elements $\langle\psi_\alpha^\text{rep}|\Psi\rangle$ and $\langle\psi_\beta^\text{rep}|\Psi\rangle$, and the temporal coupling matrix $\mathbf{T}^\text{rep}$ containing the elements $\langle\psi_\beta^\text{rep}|\frac{\mathrm{d}}{\mathrm{d} t}|\psi_\alpha^\text{rep}\rangle$.
With these definitions, equation~\eqref{eq:eom_general} can be written very compactly in the following matrix equation:
\begin{equation}
  \frac{\mathrm{d}}{\mathrm{d} t}\mathbf{c}^\text{rep}
  =
  -\left[
    \text{i}\mathbf{H}^\text{rep}
    +
    \mathbf{T}^\text{rep}
  \right]
  \mathbf{c}^\text{rep}.
  \label{eq:eom}
\end{equation}
Note that $\mathbf{T}^\text{rep}=\mathbf{v}\mathbf{K}^\text{rep}$, where $\mathbf{K}^\text{rep}$ is the non-adiabatic coupling matrix with elements $\langle\psi_\beta^\text{rep}|\nabla_\mathbf{R}|\psi_\alpha^\text{rep}\rangle$ and $\mathbf{v}$ is the nuclear velocity vector.
The basis in which matrix representations of operators are expressed are given by a superscript of the matrix.

The reasons and implications of the choice of the representation of the electronic states are discussed in the following.

\subsection*{Representations}

In a purely quantum-mechanical treatment of the nuclear motion the particular choice of the electronic basis does not influence the results, as long as the space spanned by the basis functions does not change.
In contrast, due to its semiclassical nature, the results of surface hopping are not invariant to the choice of the electronic basis,\cite{Persico2014TCA} as will be explained below.

For the purpose of this paper, it is useful to discuss three types of representations, in which $\mathbf{H}^\text{rep}$ and $\mathbf{K}^\text{rep}$ have different properties.
We start by considering the molecular Coulomb Hamiltonian (MCH), which contains only the kinetic energy of the electrons and Coulombic interactions within the molecule, but neither external fields (e.g., electric fields) nor relativistic effects (e.g., SOC):
\begin{equation}
  \hat{H}_{\text{el}}^{\text{MCH}}
  = -\sum\limits_i \left[\frac{1}{2}\nabla_i^2+\sum\limits_A \frac{Z_A}{r_{Ai}}\right]
  +\sum\limits_{i>j} \frac{1}{r_{ij}}+\sum\limits_{A>B}\frac{Z_AZ_B}{r_{AB}}.
\end{equation}
Here, the indices $i$ and $j$ run over the electrons, $A$ and $B$ over the nuclei, $Z$ is the nuclear charge, $r$ is the distance between two particles, and atomic units have been used.
The basis spanned by the (few lowest) eigenstates of the MCH Hamiltonian is denoted here as the \emph{MCH representation}.
In quantum chemistry, the MCH states are often called ``adiabatic'', but here we shall refrain from using this term since it can lead to ambiguity once SOCs are introduced.
Since the MCH is spin-independent and thus commutes with $\hat{S}^2$ and $\hat{S}_z$, the MCH states are also eigenstates of these operators; hence, they are usually labelled by their $S$ and $M_S$ values.
Furthermore, because states of different $S$ or $M_S$ values are not coupled by the MCH, generally only a single multiplicity has to be considered in the dynamics (in most cases, singlet states due to their prevalence in closed-shell organic molecules).
In the MCH representation, the matrix $\mathbf{H}^\text{MCH,MCH}$ (with matrix elements $H_{\beta\alpha}=\langle\psi_\beta^\text{MCH}|\hat{H}_{\text{el}}^\text{MCH}|\psi_\alpha^\text{MCH}\rangle$) is diagonal, but the elements of the non-adiabatic coupling matrix $\mathbf{K}^\text{MCH}$ are not zero (since the basis functions change with nuclear position) and may locally become very large.
Since nearly all electronic structure codes have been implemented to calculate MCH states, this representation is the natural one to perform on-the-fly dynamics simulations.

Let us now consider additional terms in the Hamiltonian, in particular terms which are not included in the Hamiltonian employed by standard quantum chemistry software.
These additional terms together with the MCH form the total Hamiltonian,
\begin{equation}
  \hat{H}_{\text{el}}^{\text{total}}
  =\hat{H}_{\text{el}}^{\text{MCH}}
  +\hat{H}_{\text{el}}^{\text{add}},
\end{equation}
where $\hat{H}_{\text{el}}^{\text{add}}$ represents the additional terms that are responsible for the process under investigation.
These could be, for example, the interaction between molecular dipole moments and an external electric field $-\boldsymbol{\mu}\boldsymbol{\epsilon}^{\text{ext}}$, which allows describing absorption and stimulated emission of electromagnetic radiation.
For the purpose of this contribution -- the description of ISC -- $\hat{H}_{\text{el}}^{\text{add}}$ is equal to the spin-orbit operator.
Details on the calculation of SOC are beyond of the scope of this work and can be found elsewhere.\cite{Hess1996CPL,Dyall2007,Reiher2009,Marian2012WCMS}
Upon inclusion of additional terms, the matrix representation of the total Hamiltonian in the MCH representation $\mathbf{H}^\text{total,MCH}$ (now with matrix elements $H_{\beta\alpha}=\langle\psi_\beta^\text{MCH}|\hat{H}_{\text{el}}^\text{total}|\psi_\alpha^\text{MCH}\rangle$) is not diagonal anymore, since the eigenstates of the MCH and the eigenstates of the total Hamiltonian are, in general, not identical.
Importantly, the off-diagonal couplings in $\mathbf{H}^\text{total,MCH}$ are usually delocalized over the PES -- a fact not optimal for surface hopping, since delocalized couplings can lead to a non-zero transition probability also far away from crossing regions.
Such a situation necessitates a much larger number of trajectories to sample the process correctly because surface hops may occur in a much larger phase space volume.

Besides introducing off-diagonal elements in the Hamiltonian, the spin-orbit operator lifts the degeneracy between states with the same spatial wavefunction and same $S$ but different $M_S$ (the \emph{components} of a multiplet).
As a consequence, the investigation of ISC requires to take into account all components of the relevant spin multiplets explicitly and each multiplet of spin $S$ will contribute $2S+1$ states to the model space in the simulation.
Importantly, in surface hopping simulations, the sum of the transition probabilities into all multiplet components should be independent of rotation of the molecule in the laboratory frame.
As shown by Granucci et al.,\cite{Granucci2012JCP} this requirement is not fulfilled by surface hopping in the MCH representation (including SOCs as the non-diagonal elements of $\mathbf{H}^\text{total,MCH}$), which is a problem of the MCH representation in the presence of SOCs.
In other words, if SOCs are included, the MCH basis should be regarded as sub-optimal for surface hopping.

A solution to this problem is to choose a different basis -- one in which the total Hamiltonian matrix is diagonal.
This can be achieved by a unitary transformation from the MCH basis to what we call the \emph{diagonal basis}:
\begin{equation}
  \mathbf{H}^{\text{diag}}=\mathbf{U}^\dagger\mathbf{H}^{\text{MCH}}\mathbf{U}.
  \label{eq:diagonalization}
\end{equation}
Henceforth, we exclusively discuss matrix representations of the \emph{total} Hamiltonian, so that $\mathbf{H}^\text{rep}$ always has matrix elements $H_{\beta\alpha}=\langle\psi_\beta^\text{rep}|\hat{H}_{\text{el}}^\text{total}|\psi_\alpha^\text{rep}\rangle$.

By definition, the matrix representation of the total Hamiltonian in the diagonal representation $\mathbf{H}^{\text{diag}}$ is diagonal.
\begin{equation}
  H^\text{diag}_{\beta\alpha}=\delta_{\beta\alpha}E^\text{diag}_\alpha.
\end{equation}
All couplings between the diagonal states are described by the non-adiabatic couplings $\mathbf{K}^{\text{diag}}$.
Since the non-adiabatic couplings only become large when two PES are close to each other, the requirement that all couplings are localized is fulfilled.
Additionally, the diagonalization solves the problem of rotational invariance of the multiplet components, since the eigenstates of the Hamiltonian do not depend on the molecular orientation: any rotation of the spin quantization axis could be described by a unitary transformation of the Hamiltonian $\mathbf{X}^\dagger\mathbf{H}\mathbf{X}$, which is similar to the original Hamiltonian and hence has the same eigenvalues.
Since the diagonal representation (also refered to as ''fully adiabatic'' in reference\cite{Richter2011JCTC}) solves the problems of coupling localization and rotational invariance, and also best describes the energetics of the system, it can be regarded as the optimal representation to perform surface hopping for ISC.

For completeness, we also discuss the so-called ``diabatic'' representation.\cite{Domcke2004}
This representation is defined by having time- or geometry-independent basis functions (also called ``crude adiabatic'' basis\cite{Ballhausen1972ARPC}), which means that a diabatic basis has $\mathbf{T}^\text{diab}=0$ and likewise $\mathbf{K}^\text{diab}=0$.
The coupling between diabatic states is described by off-diagonal terms in the Hamiltonian.
This basis has the advantage that no diverging non-adiabatic couplings need to be considered, and the couplings in the Hamiltonian are easier to treat numerically.
Another merit of using this electronic basis is that experimental observables are not strongly geometry-dependent and therefore, dynamical results in a diabatic basis facilitate discussion with respect to experiments.
The disadvantage of this basis is that, in general, it is not clear whether strictly diabatic states for polyatomic molecules exist\cite{Kendrick2000CPL,Baer2000CPL} and only a ``quasi-diabatic'' basis can be constructed, where most non-adiabatic couplings are small.
An additional problem of the usage of a diabatic basis is the need to first diabatize the adiabatic states obtained from quantum chemistry, a procedure which is usually not possible with on-the-fly techniques.
Hence, diabatic basis functions are in general not applicable to surface hopping simulations, at least not for full-dimensional treatments of larger molecules using the on-the-fly approach.

\begin{figure}
  \centering
  \includegraphics[scale=1.0]{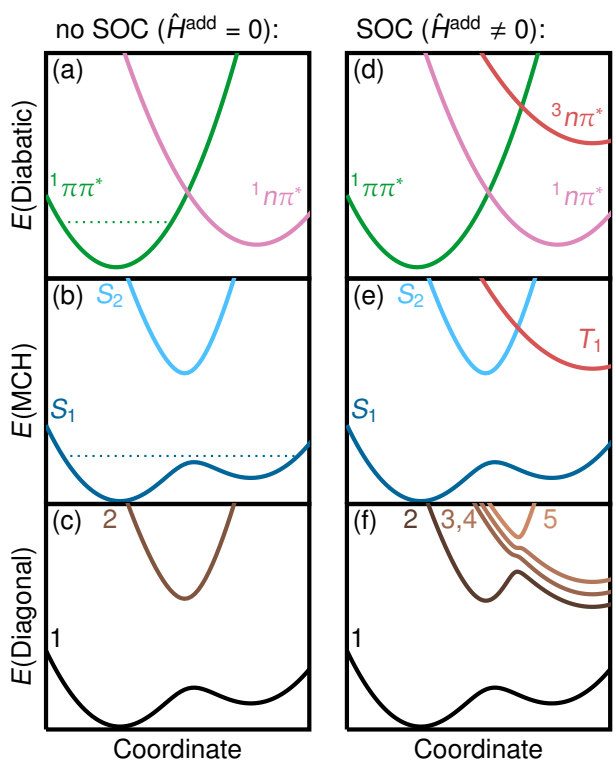}
  \caption{Examples of PES in different representations, without and with SOC. In (a) and (b), a dashed line shows the total energy available.}
  \label{fig:bases}
\end{figure}
The three electronic representations discussed before are illustrated in Figure~\ref{fig:bases}, which show examples for PES of a model system.
On the left side, a two-state system is shown.
In panel (a), two diabatic states, say $^1\pi\pi^*$ and $^1n\pi^*$, show a crossing in the center of the plot.
For the purpose of this example, we assume an arbitrary total energy (indicated by a dashed line), which is not sufficient to reach the crossing and therefore access the right potential well.
Hence, in a surface hopping simulation using these underlying potentials, the system would stay exclusively in the left minimum.
In panel (b), the same states are plotted in the MCH representation. Now the electronic wavefunction character is not preserved and hence the states are called $S_1$ and $S_2$ (by energetic ordering).
These states show a crossing strongly avoided, which lowers the energy barrier and makes accessing the right minimum classically allowed.
In this representation, a classical dynamics simulation (e.g., with surface hopping), would visit both minima.
The comparison of panels (a) and (b) clearly exemplifies the fact that the electronic representation has a direct influence on classical trajectories, which is one of the reasons why the results of surface hopping are indeed not invariant to the representation.
Panel (c) shows that in the absence of additional couplings the MCH representation is identical with the diagonal representation. The distinction between MCH and diagonal representation comes into play only if $\hat{H}_{\text{el}}^{\text{add}}\neq 0$.

On the right side of figure~\ref{fig:bases}, we show an exemplary three-state system (two singlets and one triplet), which are coupled by SOC.
In panel (d), the diabatic PES of the $^1\pi\pi^*$ and $^1n\pi^*$ are unchanged compared to (a), since the SOCs only appear as off-diagonal elements of the Hamiltonian.
The same holds true for panel (e), which is identical to (b) except for the addition of the triplet state.
Only when transforming from the MCH to the diagonal representation (panel (f)) the PES are affected by SOC.
Importantly, SOC lifts the degeneracy of the triplet components, giving rise to five states in total.
A trajectory will explicitly depend on which triplet component it is moving in, again showing that the choice of the representation directly influences the result of classical trajectories, and that the diagonal representation performs best in this regard.
Together with the fact that the diagonal representation optimally localizes all couplings in $\mathbf{H}^\text{rep}$ and $\mathbf{K}^\text{rep}$, we conclude that this representation is indeed optimal for surface hopping.

One should note, however, that the validity of other representations depends crucially on the strength of the coupling between the electronic states.
If the SOCs are large, a large effect on the form of the PES can be expected and the diagonal representation is mandatory.
For very weak SOC, the PES are only weakly deformed and other representations should yield approximately the same results, which justifies the usage of, e.g., the MCH representation in some forms of surface hopping including additional couplings.\cite{Gai1992JAP,Mitric2009PRA,Carbogno2010PRB,Cui2014JCP}

\subsection*{Surface hopping in the diagonal representation}

Currently there is no quantum chemistry program able to calculate all necessary properties needed for surface hopping in the diagonal basis.
While gradients of diagonal states including SOC are already described for semi-empirical wavefunctions,\cite{Granucci2011JCC} no such gradients are available up to now for ab initio electronic structure methods and neither are the corresponding non-adiabatic couplings.
Furthermore, quantum chemistry in the diagonal basis including SOCs (see references~\cite{Dyall2007,Reiher2009,Mai2014JCP_reindex} and references therein) is in many cases significantly more expensive than quantum chemistry in the MCH basis. 
The higher cost is related to the fact that the spin-orbit Hamiltonian does not commute with $\hat{S}^2$ or $\hat{S}_z$ and hence it is not block-diagonal.
Additionally, more demanding basis sets are usually required\cite{Dyall2007,Reiher2009} for quantum chemical calculations in the diagonal basis than for MCH ones.

As a consequence, quantum chemistry in the MCH basis is nowadays the most advanced one.
A number of quantum chemistry software suites have efficient implementations of analytical gradients, non-adiabatic couplings and SOC matrix elements,\cite{Lischka2012,Werner2012,Aquilante2010JCC} which can be employed for the on-the-fly calculations of surface hopping.
Hence, a pragmatical approach for trajectory surface hopping methods is to combine quantum chemical data in the MCH representation with surface hopping in the diagonal representation.
This is the essence of the ab initio molecular dynamics methodology \textsc{Sharc},\cite{Richter2011JCTC} recently developed in our group.
The basic idea is to construct the matrix representation of the total Hamiltonian in a \emph{small set of relevant MCH states} (the model space) and diagonalize the Hamiltonian only in this small basis.
The SOCs between the model space states and all remaining states are neglected, making this a form of quasi-degenerate perturbation theory (``QDPT'', see, e.g., reference~\cite{Mai2014JCP_reindex}).
The obtained eigenstates are approximations to the true eigenstates of $\hat{H}_\text{el}^\text{total}$, but for small SOC this is a good approximation.

In the following, we describe how to perform surface hopping in a diagonal basis obtained from QDPT, as it is implemented in the \textsc{Sharc} code.\cite{Mai2014SHARC}

\subsubsection*{Equation of motion}

To make surface hopping in the diagonal representation practical, one needs a suitable equation of motion for the diagonal coefficients $\mathbf{c}^\text{diag}$ and a corresponding expression for the hopping probabilities $\mathbf{h}^\text{diag}$ based on the MCH quantities $\mathbf{H}^\text{MCH}$ and $\mathbf{K}^\text{MCH}$.
Additionally, one needs to construct the diagonal gradient $-\nabla E_\beta^\text{diag}$ from the MCH gradients.

We start with the equation of motion~\eqref{eq:eom} in the MCH basis:
\begin{equation}
  \frac{\mathrm{d}}{\mathrm{d} t}\mathbf{c}^{\text{MCH}}=
  -\left[
    \text{i}\mathbf{H}^{\text{MCH}}
    +\mathbf{v}\mathbf{K}^{\text{MCH}}
  \right]
  \mathbf{c}^{\text{MCH}}.\label{eq:eom_MCH}
\end{equation}
The diagonal wavefunction coefficients $\mathbf{c}^\text{diag}$ can be easily obtained from $\mathbf{c}^\text{MCH}$, using the transformation matrix $\mathbf{U}$:
\begin{equation}
  \mathbf{c}^\text{diag}=\mathbf{U}^\dagger\mathbf{c}^\text{MCH}.
  \label{eq:c_trans}
\end{equation}
Inserting the latter into equation (\ref{eq:eom_MCH}) and pre-multiplying with $\mathbf{U}^\dagger$, we obtain:
\begin{equation}
  \frac{\mathrm{d}}{\mathrm{d} t}\mathbf{c}^{\text{diag}}=
  -\mathbf{U}^\dagger\left[
    \text{i}\mathbf{H}^{\text{MCH}}
    +\mathbf{v}\mathbf{K}^{\text{MCH}}
  \right]\mathbf{U}
  \mathbf{c}^{\text{diag}}
  -\mathbf{U}^\dagger\frac{\mathrm{d} \mathbf{U}}{\mathrm{d} t}
  \mathbf{c}^{\text{diag}},\label{eq:eom_diag}
\end{equation}
Note how the last term on the right-hand side arises since the transformation matrix $\mathbf{U}$ implicitly depends on time via the time dependence of the Hamiltonian.
Given a coefficient vector $\mathbf{c}^\text{diag}(t)$ and the matrices $\mathbf{H}^\text{MCH}$ and $\mathbf{K}^\text{MCH}$, equation~\eqref{eq:eom_diag} can be integrated using standard Runge-Kutta-type numerical methods, or using the exponential operator method for sufficiently short time steps:
\begin{equation}
  \mathbf{c}^{\text{diag}}(t+\Delta t)=
  \exp\left[
    -
    \left(
      \text{i}\mathbf{U}^\dagger\mathbf{H}^{\text{MCH}}\mathbf{U}
      +\mathbf{v}\mathbf{U}^\dagger\mathbf{K}^{\text{MCH}}\mathbf{U}
      +\mathbf{U}^\dagger\frac{\mathrm{d} \mathbf{U}}{\mathrm{d} t}
    \right)\Delta t
  \right]
  \mathbf{c}^{\text{diag}}(t),
  \label{eq:one-step-prop}
\end{equation}
where we omitted the explicit time-dependence of $\mathbf{U}$, $\mathbf{H}^\text{MCH}$ and $\mathbf{K}^\text{MCH}$ for brevity.
However, in this scheme the term $\mathbf{U}^\dagger\frac{\mathrm{d} \mathbf{U}}{\mathrm{d} t}$ leads to serious numerical difficulties.
First, the unitary matrix $\mathbf{U}$ is not uniquely defined by equation~\eqref{eq:diagonalization} -- each eigenvector could be multiplied by a complex phase factor $\text{e}^{\text{i}\phi}$ and still be an eigenvector to the same eigenvalue.
This non-uniqueness makes it impossible to calculate $\mathbf{U}^\dagger\frac{\mathrm{d} \mathbf{U}}{\mathrm{d} t}$ from finite differences, because $\mathbf{U}$ is not continuous.
Second, $\mathbf{U}^\dagger\frac{\mathrm{d} \mathbf{U}}{\mathrm{d} t}$ may become large if two MCH states cross and the norm of their mutual coupling tends to zero, which can generally occur in all systems.
Using shorter time steps $\Delta t$ alleviates the problem, but the numerical precision of usual floating point calculations limits the size of the time step.
Furthermore, if the mutual coupling is zero, $\mathbf{U}^\dagger\frac{\mathrm{d} \mathbf{U}}{\mathrm{d} t}$ diverges.

These problems can be easily circumvented by integrating (for small $\Delta t$) the equation of motion in the MCH basis~\eqref{eq:eom_MCH},
\begin{equation}
  \mathbf{c}^{\text{MCH}}(t+\Delta t)=
  \exp\left[
    -
    \left(
      \text{i}\mathbf{H}^{\text{MCH}}
      +\mathbf{v}\mathbf{K}^{\text{MCH}}
    \right)\Delta t
  \right]
  \mathbf{c}^{\text{MCH}}(t),
\end{equation}
and only then inserting equation~\eqref{eq:c_trans}:
\begin{equation}
  \mathbf{c}^{\text{diag}}(t+\Delta t)=
  \mathbf{U}^\dagger(t+\Delta t)
  \exp\left[
    -
    \left(
      \text{i}\mathbf{H}^{\text{MCH}}
      +\mathbf{v}\mathbf{K}^{\text{MCH}}
    \right)\Delta t
  \right]
  \mathbf{U}(t)
  \mathbf{c}^{\text{diag}}(t).
  \label{eq:three-step-prop}
\end{equation}
In this way, one obtains a propagation equation which is fully equivalent to~\eqref{eq:one-step-prop}, but where $\mathbf{U}^\dagger\frac{\mathrm{d} \mathbf{U}}{\mathrm{d} t}$ is not explicitly involved.
Note that this equation describes a propagation which proceeds in three distinct steps:
\begin{enumerate}
  \item transforming $\mathbf{c}^{\text{diag}}(t)$ to $\mathbf{c}^{\text{MCH}}(t)$,
  \item propagating $\mathbf{c}^{\text{MCH}}(t)$ to $\mathbf{c}^{\text{MCH}}(t+\Delta t)$, and
  \item transforming $\mathbf{c}^{\text{MCH}}(t+\Delta t)$ to $\mathbf{c}^{\text{diag}}(t+\Delta t)$.
\end{enumerate}
We call this approach the \emph{three-step propagator}, and write it in short as
\begin{equation}
  \mathbf{c}^{\text{diag}}(t+\Delta t)=
  \underbrace{
  \mathbf{U}^\dagger(t+\Delta t)
  \mathbf{P}^\text{MCH}(t+\Delta t,t)
  \mathbf{U}(t)
  }_{\mathbf{P}^\text{diag}(t+\Delta t,t)}
  \mathbf{c}^{\text{diag}}(t),
  \label{eq:three-step-prop-brief}
\end{equation}
where $\mathbf{P}^\text{MCH}(t+\Delta t,t)$ is the propagator matrix in the MCH basis, which acts on the MCH coefficients at time $t$ to propagate it to time $t+\Delta t$.
We note here that $\mathbf{P}^\text{MCH}(t+\Delta t,t)$ could also be calculated differently from equation~\eqref{eq:three-step-prop}, for example using the local diabatization procedure of Granucci et al.\cite{Granucci2001JCP,Plasser2012JCP}
The condition that the time step $\Delta t$ is short enough can be fulfilled by choosing a shorter time step for the electronic propagation than for the nuclear propagation and linearly interpolating the electronic properties (see Appendix A for the corresponding equations).

The three-step-propagator methodology has a number of advantageous properties.
First, it accomplishes our main goal, i.e., formulating the propagation of the diagonal coefficients based on MCH quantities.
Second, it is unitary (the matrix exponential of an anti-Hermitian matrix is always unitary), and thus the wavefunction norm is automatically conserved during the propagation.
And third, the numerical computation of $\mathbf{U}^\dagger\frac{\mathrm{d} \mathbf{U}}{\mathrm{d} t}$ is avoided, providing a much more stable propagation with respect to the time step and the involved couplings in the Hamiltonian.

The following example shows the numerical superiority of the three-step propagator over the more na\"ive single-step method in equation~\eqref{eq:one-step-prop}.
We consider a simple one-dimensional system consisting of two harmonic oscillators coupled by a constant off-diagonal term $\xi$ (atomic units):
\begin{equation}
  \mathbf{H}^\text{MCH}(x)=
  \begin{pmatrix}
    0.1x^2 &\xi\\
    \xi      &0.1(x-2)^2\\
  \end{pmatrix}.
\end{equation}
The initial position of the system is set at $x=10$ in the upper diagonal state and the reduced mass is $m=0.2$.
We simulate a single pass of a trajectory through the crossing region and record the total change in the electronic populations depending on the strength of $\xi$.
We performed the simulations with equation~\eqref{eq:three-step-prop} and a time step of $10^{-2}$~fs.
For comparison we also used equation~\eqref{eq:one-step-prop}, with different time steps between $10^{-4}$ and $10^{-8}$~fs.
In order to show that controlling the complex phases of the columns of $\mathbf{U}$ is imperative when using equation~\eqref{eq:one-step-prop}, results with and without phase control (using the algorithm outlined in Appendix B) are shown.
Figure~\ref{fig:ex_1} shows the population of the initially empty state ($|c^\text{MCH}_1(0)|^2=0$) after passing the crossing.
According to Landau-Zener theory, for sufficiently small couplings, the population after the crossing $|c^\text{MCH}_1(t>t_\text{cross})|^2$ should be proportional to $\xi^2$.
\begin{figure}
  \centering
  \includegraphics[scale=1.0]{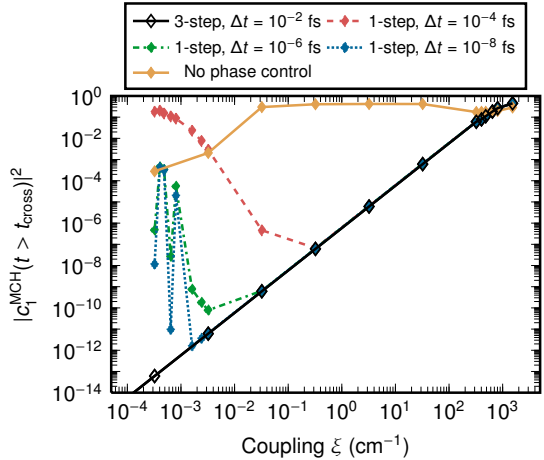}
  \caption{Example showing how the population transfer depends on the strength of SOC and on the propagation time substep.}
  \label{fig:ex_1}
\end{figure}
The three-step propagator shows the expected proportionality of $|c^\text{MCH}_1(t>t_\text{cross})|^2$ and $\xi^2$, even though a comparatively long time step was employed.
Even if $\xi$ is set equal to zero the three-step propagator delivers the correct result of zero population transfer.
The one-step propagator data shows that equation~\eqref{eq:one-step-prop} only performs well when the coupling $\xi$ is large, while for smaller couplings, population transfer is systematically overestimated.
The use of comparatively smaller time steps helps in the case of small couplings, but even the shortest time step of $10^{-8}$~fs fails for $\xi<10^{-3}$ cm$^{-1}$ when using the one-step propagator.
When $\xi$ tends to zero, equation~\eqref{eq:one-step-prop} predicts that the population $|c^\text{MCH}_1(t>t_\text{cross})|^2$ tends to one, which is of course unphysical.
Not controlling the phase of $\mathbf{U}$ leads to a completely erratic behaviour and a much too large population transfer, which is basically independent from $\xi$.

The message from this figure is that the one-step propagator is very dangerous to use in the presence of very small couplings.
This is a very serious problem in ISC simulations because, even though a limited number of SOC matrix elements may be large, the problematic small SOC matrix elements are generally present in all systems. 
Especially in small molecules, the selection rules of SOC render some matrix elements equal to zero, and the El-Sayed rule\cite{El-Sayed1963JCP} predicts that SOCs between states with very similar wavefunction character are small.
Therefore, the use of the three-step propagator method as outlined above is recommended to avoid the problems in the wavefunction propagation in such situations.

\subsubsection*{Hopping probabilities}

The use of the matrix exponential for the propagation of the coefficients is not compatible with the original expression for the hopping probabilities given by Tully.\cite{Tully1990JCP}
In the diagonal representation, this expression contains $\frac{\mathrm{d}}{\mathrm{d} t}\mathbf{c}^\text{diag}$, which according to equation~\ref{eq:eom_diag} explicitly depends on $\mathbf{U}^\dagger\frac{\mathrm{d} \mathbf{U}}{\mathrm{d} t}$, giving rise to the numerical problems mentioned above.
Consequently, it is necessary to obtain a different expression for the hopping probabilities $h_{\beta\rightarrow\alpha}$ -- yet based on the fewest switches criterion.
Here we use a variant of the equation derived in reference~\cite{Granucci2001JCP}, where the hopping probability $h_{\beta\rightarrow\alpha}$ is just a function of the old and new coefficients and the propagator matrix:
\begin{equation}
  h_{\beta\rightarrow\alpha}^\text{diag}=
  \left(
    1-
    \frac{
      \left|
        c_\beta^{\text{diag}}(t+\Delta t)
      \right|^2
    }{
      \left|
        c_\beta^{\text{diag}}(t)
      \right|^2
    }\right)
    \times
    \frac{
      \Re\left[
        c^{\text{diag}}_\alpha(t+\Delta t)
        \left(P^\text{diag}_{\alpha\beta}\right)^*
        \left(
          c^{\text{diag}}_\beta(t)
        \right)^*
      \right]
    }{
      \left|
        c^{\text{diag}}_\beta(t)
      \right|^2
      -\Re\left[
        c^{\text{diag}}_\beta(t+\Delta t)
        \left(P^\text{diag}_{\beta\beta}\right)^*
        \left(
          c^{\text{diag}}_\beta(t)
        \right)^*
      \right]
    }.
  \label{eq:probabilities}
\end{equation}
Negative hopping probabilities are set to zero, as is $h_{\beta\rightarrow\beta}$.
The equation can be derived similarly as in reference\cite{Granucci2001JCP}, but using $\mathbf{P}^\dagger\mathbf{c}(t+\Delta t)=\mathbf{c}(t)$ instead of $\mathbf{c}(t+\Delta t)=\mathbf{P}\mathbf{c}(t)$.
One advantage of equation~\eqref{eq:probabilities} over the one in reference\cite{Granucci2001JCP} is that the denominator in equation~\eqref{eq:probabilities} does not tend to zero when the population transfer becomes large, giving more accurate results. 
Moreover, equation~\eqref{eq:probabilities} has the nice property that the hopping probabilities are not affected by the non-uniqueness of the transformation matrix $\mathbf{U}$.
Hence, no phase adjustment between $\mathbf{U}(t)$ and $\mathbf{U}(t+\Delta t)$ is necessary at all in the \textsc{Sharc} methodology.

After the propagation for a particular time step has been performed, equation~\eqref{eq:probabilities} can be directly used to calculate hopping probabilities.
To choose the active state for the next time step, a random number $r$ from the interval $0 \leq r \leq 1$ is selected.
A hop to state $\alpha$ is performed, if
\begin{equation}
  \sum\limits_{i=1}^{\alpha-1} h_{\beta\rightarrow i} < r \le \sum\limits_{i=1}^{\alpha-1} h_{\beta\rightarrow i}+h_{\beta\rightarrow\alpha}
  \label{eq:decision}
\end{equation}
i.e., if the random number $r$ lies in an interval with a width proportional to the hopping probability.

\subsubsection*{Gradients}

Equations~\eqref{eq:diagonalization}, \eqref{eq:three-step-prop} and \eqref{eq:probabilities} define all that is needed to perform the electronic part of the surface hopping algorithm, namely to determine the active state for the next time step.
The remaining necessary element of an algorithm for surface hopping in the diagonal representation is the construction of the gradient of the diagonal states.

We start with the derivative of a matrix element of the total Hamiltonian in the diagonal basis:
\begin{equation}
  \nabla_\mathbf{R} \left\langle\psi_\beta^\text{diag}\middle|\hat{H}_\text{el}^\text{total}\middle|\psi_\alpha^\text{diag}\right\rangle
  =
  \nabla_\mathbf{R}H_{\beta\beta}^\text{diag}\delta_{\beta\alpha}.
  \label{eq:deriv1}
\end{equation}
By applying the product rule to the left-hand side and using the fact that the $|\psi^\text{diag}\rangle$ are the eigenfunctions of $\hat{H}_\text{el}^\text{total}$, equation~\eqref{eq:deriv1} can be rearranged to\cite{Doltsinis2002JTCC}
\begin{equation}
  \left\langle\psi_\beta^\text{diag}\middle|\nabla_\mathbf{R}\hat{H}_\text{el}^\text{total}\middle|\psi_\alpha^\text{diag}\right\rangle
  =
  \nabla_\mathbf{R}H_{\beta\beta}^\text{diag}\delta_{\beta\alpha}
  -(H_{\beta\beta}^\text{diag}-H_{\alpha\alpha}^\text{diag})\mathbf{K}_{\beta\alpha}^\text{diag},
  \label{eq:deriv2}
\end{equation}
where $H_{\beta\beta}^\text{diag}$ and $H_{\alpha\alpha}^\text{diag}$ are eigenvalues of the total Hamiltonian and $\mathbf{K}_{\beta\alpha}^\text{diag}$ is a matrix element of the non-adiabatic coupling matrix $\mathbf{K}^\text{diag}$ (note that the elements of $\mathbf{K}^\text{diag}$ are itself vectors in the space of the nuclear distortions).
Importantly, this equation contains $\nabla_\mathbf{R}H_{\beta\beta}^\text{diag}$, which is the diagonal gradient necessary for the dynamics.

The left-hand side of equation~\eqref{eq:deriv2} can be written in terms of the MCH states and the transformation matrix $\mathbf{U}$ as
\begin{equation}
  \left\langle\psi_\beta^\text{diag}\middle|\nabla_\mathbf{R}\hat{H}_\text{el}^\text{total}\middle|\psi_\alpha^\text{diag}\right\rangle
  =
  \sum\limits_i\sum\limits_j
  U^*_{i\beta}U_{j\alpha}
  \left\langle\psi_i^\text{MCH}\middle|\nabla_\mathbf{R}(\hat{H}_\text{el}^\text{MCH}+\hat{H}_\text{el}^\text{add})\middle|\psi_j^\text{MCH}\right\rangle.
  \label{eq:deriv3}
\end{equation}
The critical term in this equation is the derivative of the additional Hamiltonian terms with respect to the nuclear displacements $\nabla_\mathbf{R}\hat{H}_\text{el}^\text{add}$.
If -- as in the present case -- $\hat{H}_\text{el}^\text{add}$ is a spin-orbit operator, then this derivative is (up to our knowledge) not available from standard ab initio software.
However, due to the short-range nature of the spin-orbit interaction,\cite{Albrecht1963JCP} the derivative $\nabla_\mathbf{R}\hat{H}_\text{el}^\text{add}$ is expected to be small, and is hence ignored.
The remainder of equation~\eqref{eq:deriv3} can thus be written as
\begin{equation}
  \left\langle\psi_\beta^\text{diag}\middle|\nabla_\mathbf{R}\hat{H}_\text{el}^\text{total}\middle|\psi_\alpha^\text{diag}\right\rangle
  =
  \sum\limits_i\sum\limits_j
  U^*_{i\beta}U_{j\alpha}
  \left\langle\psi_i^\text{MCH}\middle|\nabla_\mathbf{R}\hat{H}_\text{el}^\text{MCH}\middle|\psi_j^\text{MCH}\right\rangle.
  \label{eq:deriv4}
\end{equation}
As the $\psi_i^\text{MCH}$ are the eigenfunctions of $\hat{H}_\text{el}^\text{MCH}$, we can use an analogue form of equation~\eqref{eq:deriv2} to arrive at
\begin{equation}
  \nabla_\mathbf{R}H_\beta^\text{diag}\delta_{\beta\alpha}-(H_{\beta\beta}^\text{diag}-H_{\alpha\alpha}^\text{diag})\mathbf{K}_{\beta\alpha}^\text{diag}
  =
  \sum\limits_i\sum\limits_j
  U^*_{i\beta}U_{j\alpha}
  \left[\nabla_\mathbf{R}H_i^\text{MCH}\delta_{ij}-(H_{ii}^\text{MCH}-H_{jj}^\text{MCH})\mathbf{K}_{ij}^\text{MCH}\right].
\end{equation}
If we define the general gradient matrix in an arbitrary representation $\mathbf{G}^\text{rep}$ with matrix elements
\begin{equation}
  G^\text{rep}_{ij}
  =
  \nabla_\mathbf{R}H_{ii}^\text{rep}\delta_{ij}-(H_{ii}^\text{rep}-H_{jj}^\text{rep})\mathbf{K}_{ij}^\text{rep},
  \label{eq:grad:def}
\end{equation}
then we can write the above equation compactly as
\begin{equation}
  \mathbf{G}^\text{diag}=
  \mathbf{U}^\dagger\mathbf{G}^{\text{MCH}}\mathbf{U}.
  \label{eq:grad:trans}
\end{equation}

Equations~\eqref{eq:grad:def} and \eqref{eq:grad:trans} give a rule how to obtain the diagonal gradients and non-adiabatic couplings from the MCH gradients and non-adiabatic couplings.
From an algorithmic point of view, the diagonal vector properties can be obtained in the following steps:
\begin{enumerate}
  \item Obtain the energies $H_{ii}^\text{MCH}$, gradients $\nabla_\mathbf{R}H_{ii}^\text{MCH}$ and non-adiabatic couplings $\mathbf{K}_{ij}^\text{MCH}$ from quantum chemistry.
  \item Construct $\mathbf{G}^{\text{MCH}}$ from these quantities using equation~\eqref{eq:grad:def}.
  \item Transform $\mathbf{G}^{\text{MCH}}$ into the diagonal basis using equation~\eqref{eq:grad:trans}.
  \item Obtain the diagonal gradients $\nabla_\mathbf{R}H_{\beta\beta}^\text{diag}=\mathbf{G}^\text{diag}_{\beta\beta}$.
  \item If needed, obtain the non-adiabatic coupling vectors in the diagonal basis $\mathbf{K}_{\beta\alpha}^\text{diag}=(H_{\beta\beta}^\text{diag}-H_{\alpha\alpha}^\text{diag})^{-1}\mathbf{G}^\text{diag}_{\beta\alpha}$.
\end{enumerate}

As one can see from the last paragraphs, the gradients in the diagonal basis are linear combinations of the MCH gradients and the non-adiabatic coupling vectors.
Therefore, even if only a single diagonal gradient is needed, several MCH gradients and coupling vectors need to be obtained from quantum chemistry.
Compared to dynamics in the MCH basis, where only a single gradient needs to be computed at each time step, for dynamics in the diagonal basis, in principle, all gradients are required, adding computational cost to the simulation.

\begin{figure}
  \centering
  \includegraphics[scale=1.0]{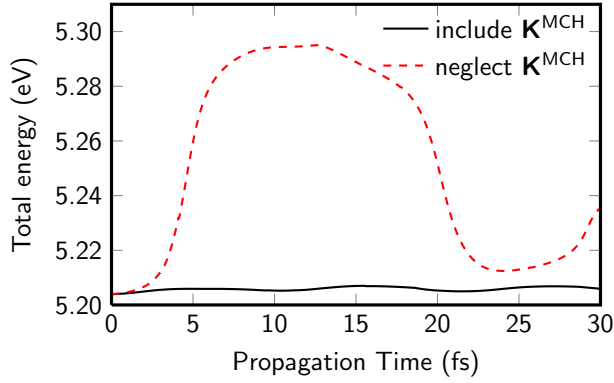}
  \caption{Example showing how neglecting non-adiabatic couplings in the gradient transformation affects the total energy conservation.}
  \label{fig:ex_2}
\end{figure}

As a consequence of the above-said, the surface hopping procedure needs the non-adiabatic coupling vectors for two distinct steps: the wavefunction propagation and the gradient transformation.
Unfortunately, the calculation of non-adiabatic couplings is presently only available in some quantum chemistry codes. 
If non-adiabatic couplings are not available, it is still possible to calculate the propagator matrix $\mathbf{P}$ using other schemes, see, e.g., references~\cite{Granucci2001JCP,Hammes-Schiffer1994JCP}.
However, for the gradient transformation no alternative schemes without non-adiabatic couplings are available, and, as an approximation $\mathbf{K}_{ij}^\text{MCH}$ has to be neglected in equations~\eqref{eq:grad:def} and \eqref{eq:grad:trans}.
This introduces an error which is largest near weakly-avoided crossings (precisely where $\mathbf{K}_{ij}^\text{MCH}$ is large).
Figure~\ref{fig:ex_2} shows the total energy of two exemplary trajectories,\footnote{The simulations were performed on seleno-acrolein at the CASSCF(6,5)/ano-rcc-vdzp level of theory including two singlet and two triplet electronic states.
The initial geometry was the ground state minimum geometry and the initial state the $S_1$.
The initial velocities of all atoms were zero, except for the terminal hydrogen atoms, which had velocities of 0.008~a.u.\ out of the molecular plane and in opposite directions, thus giving some initial momentum for torsion around the C=C double bond.\cite{Mai2014JCP_reindex}} one calculated with equations~\eqref{eq:grad:def} and \eqref{eq:grad:trans} and another where $\mathbf{K}_{ij}^\text{MCH}$ is neglected.
As it can be seen in this extreme case, neglecting the contribution of the MCH non-adiabatic coupling vectors to the diagonal gradients leads to a violation of energy conservation already after a few fs.
Hence, whenever possible it is recommended to include the non-adiabatic couplings in the gradient calculations or, if these are not included, one should check the total energy conservation carefully. 

\subsubsection*{An algorithm for surface hopping in the diagonal representation}

To summarize, given some initial geometry $\mathbf{R}(t)$, velocity $\mathbf{v}(t)$, acceleration $\mathbf{a}(t)$, coefficients $\mathbf{c}(t)$ and active state $\beta(t)$, the following steps should be performed to propagate the dynamics to time $t+\Delta t$:
\begin{enumerate}
  \item Use equations~\eqref{eq:vv:a1} and \eqref{eq:vv:r} to find the new geometry $\mathbf{R}(t+\Delta t)$.
  \item Obtain $\mathbf{H}^\text{MCH}$, $\mathbf{K}^\text{MCH}$ and the gradients $-\nabla H^\text{MCH}_{ii}$ from quantum chemistry software.
  \item Find $\mathbf{U}$ from equation~\eqref{eq:diagonalization}.
  \item Use equations~\eqref{eq:grad:def} and \eqref{eq:grad:trans} to find $-\nabla H^\text{diag}_{\beta\beta}(t+\Delta t)$.
  \item Use equations~\eqref{eq:vv:a2} and \eqref{eq:vv:v} to find the new velocity $\mathbf{v}(t+\Delta t)$.
  \item Use equation~\eqref{eq:three-step-prop} to find the propagator $\mathbf{P}(t+\Delta t,t)$ and the new coefficients $\mathbf{c}(t+\Delta t)$.
  \item Use equation~\eqref{eq:probabilities} to find the hopping probabilities $h_{\beta\rightarrow\alpha}$ and equation~\eqref{eq:decision} to find the new active state $\beta(t+\Delta t)$.
  \item If a hop occured, ensure total energy conservation by adjusting $\mathbf{v}(t+\Delta t)$ as in regular surface hopping (see, e.g., references~\cite{Tully1971JCP,Fabiano2008CP}).
  \item Apply some form of decoherence correction\cite{Zhu2004JCP,Granucci2007JCP,Granucci2010JCP} to the diagonal coefficients $\mathbf{c}(t+\Delta t)$.
\end{enumerate}

A few comments regarding the different electronic state representations encountered in this workflow are appropriate at this point.
First, it should be mentioned that using MCH quantum chemistry data to perform dynamics in the diagonal basis is actually an approximation based on a QDPT treatment and hence neglects the coupling to states outside the model space.
Moreover, the fully consistent diagonal gradients could only be calculated if the derivatives of the SOCs were known.
However, in view of the current state-of-the-art of quantum chemistry, dynamics directly in the diagonal basis is not possible because the necessary properties are not available.
Second, while the diagonal representation is optimal for surface hopping dynamics, it is not optimal for the interpretation of the results, since the character and spin of the diagonal states is geometry-dependent.
Therefore, it is usually advantageous to transform the results (e.g., populations) back into the MCH basis, where the states can be characterized by their $S$ and $M_s$ quantum numbers.
This facilitates, for example, the distinction between IC and ISC processes, which are otherwise equivalent in the diagonal picture.
Of course, this \emph{a posteriori} transformation to MCH states is reasonable only for systems with SOC small enough so that the $LS$ coupling scheme is valid.
Third, a transformation of the results from the MCH into a ``spectroscopic'' basis (an approximate diabatic basis where spectroscopic observables depend strongly on the state, but weakly on geometry, e.g., states of $\pi\pi^*$ and $n\pi^*$ character) is even more convenient than the MCH basis to facilitate the interpretation of the results and is recommended as long as an \emph{a posteriori} diabatization scheme can be devised for the system.


\section*{\sffamily \Large Conclusion and Outlook}

In this contribution it has been shown how spin-orbit coupling (SOC) can be treated on the same footing as non-adiabatic coupling in semi-classical surface hopping trajectory methods.
To this aim, a general and accurate approach relying on the diagonalization of the electronic Hamiltonian including all potential couplings has been described.
This methodology was coined ``surface hopping including arbitrary couplings'' (\textsc{Sharc}) in our original publication\cite{Richter2011JCTC} because it is not limited to SOC, but can treat all kinds of couplings in the same fashion.
An on-the-fly implementation of this methodology is available in the \textsc{Sharc} molecular dynamics suite.\cite{Mai2014SHARC}

While the \textsc{Sharc} methodology is generally applicable to very large systems (due to the classical nuclear dynamics approach), the size of the systems for which ISC can be studied is actually limited by the availability of suitable quantum chemistry methods.
These methods need to provide accurate excitation energies, analytical gradients, SOC matrix elements, and non-adiabatic couplings (or equivalent properties like overlap matrices\cite{Granucci2001JCP,Plasser2012JCP}) for several excited states at once.
If laser interactions are considered in the framework of \textsc{Sharc} (few applications are found in the literature so far\cite{Richter2011JCTC,Marquetand2011FD,Bajo2011JPCA,Bajo2014JCP}), then the quantum chemistry also needs to provide transition dipole moments.
Additionally, the quantum chemical methods should be able to give a balanced and accurate description of all states over the whole potential energy surface, which usually requires multi-configurational wavefunctions.
These requirements are currently fulfilled for several implementations\cite{Aquilante2010JCC,Werner2012,Lischka2012} of the multi-configurational self-consistent field (MCSCF) method\cite{Bearpark2007JPPA,Olsen2011IJQC} and for multi-reference configuration interaction (MRCI) as implemented in \textsc{Columbus}.\cite{Lischka2012}
Using MCSCF or MRCI, already a number of ISC dynamics simulations have been reported using \textsc{Sharc}.\cite{Richter2012JPCL,Richter2014PCCP,Mai2013C,Mai2014JCP_SO2,Martinez-Fernandez2014JCTC,Corrales2014PCCP}

The progress of general ab initio molecular dynamics schemes like \textsc{Sharc} strongly depends on the development of electronic structure methods that can go beyond MCSCF, which is one of the most popular methods for on-the-fly excited-state dynamics.
A obvious step further is complete-active-space perturbation theory (e.g., CASPT2\cite{Pulay2011IJQC}), which offers very good accuracy (oftentimes better than MRCI) and performance (usually much faster than MRCI).
Few surface hopping studies based on CASPT2 have been reported already,\cite{Mori2009CPL,Kuhlman2012FD,Mori2012JPCA,Nakayama2013JCP} albeit no ISC dynamics simulation has been published until now, since no current implementation of CASPT2 can provide all necessary quantities.
Other promising electronic structure methods are, e.g., the approximately size-consistent MRCI variant linear-response-theory multi-reference average-quadratic coupled-cluster (LRT-MRAQCC)\cite{Mai2014JCP_reindex} or density matrix normalization group (DMRG) approaches.\cite{Marti2010ZPC}
Furthermore, it will be possible to employ time-dependent density functional theory (TDDFT) in ISC dynamics studies with the help of recent developments,\cite{Curchod2013C,FrancodeCarvalho2014JCP} although care has to be taken that the excited-state potential energy surfaces are described correctly in the case of strongly multi-configurational state character.

Semi-empirical electronic structure methods for excited states\cite{Thiel2014WCMS} are very efficient in comparison to ab initio or TDDFT methods and can be applied as well to ISC dynamics simulations.
There is already an implementation equivalent to \textsc{Sharc},\cite{Granucci2012JCP} which has been applied to study ISC dynamics on thioguanine.\cite{Martinez-Fernandez2014CS}
Additionally, semi-empirical calculations of ISC dynamics already have been carried out using a conventional surface hopping approach,\cite{Cui2014JCP} i.e., without the advantages provided by the \textsc{Sharc} methodology as outlined in this contribution.
In any case, the availability of TDDFT and semi-empirical methods for surface hopping will open wide opportunities to describe larger molecules or achieve longer propagation times than it is currently feasible with multi-configurational wavefunction methods.

A further challenge in trajectory surface methods is to include an efficient and accurate description of solvents and complex molecular environments in the dynamics.
Considerable effort in this direction has already been made for standard surface hopping using mixed quantum-mechanics/molecular-mechanics (QM/MM) methods, in which the main chromophore is treated quantum-chemically and the environment with force fields.\cite{Persico2003JMS-T,Senn2009ACIE,Hayashi2009BJ,Fingerhut2012JCP}

Finally, in order to relate the computational results to experimental data, not only the excited-state dynamics has to be simulated, but also the subsequent processes occuring during the experimental probing, i.e., it is necessary to actually simulate the experimental observables.
A typical example is transient absorption spectroscopy,\cite{Pollard1992ARPC} which can be simulated by including highly excited states and the corresponding transition dipole moments in the calculations.
However, an accurate description of these highly excited states is far from trivial.
Another typical probing technique is based on ionization of the studied molecule, yielding for example time-dependent photo-electron spectra.\cite{Stolow2003ARPC}
In order to simulate such ionizations, it is necessary to describe the wavefunctions of the neutral system, the ionized species as well as the outgoing electron; from these wavefunctions, the corresponding wavefunction overlaps as well as the associated transition dipole moments need to be calculated.
This is quite a formidable task that has not yet been achieved in full, so that a detailed understanding of molecular ionization is still out of reach.
Therefore, further developments are necessary to achieve the ultimate goal of simulating all the processes involved in actual experiments from the beginning to the end. Yet, given the increasing interest in dynamical processes including ISC, it is expected that the coming years will see an increasing number of surface hopping trajectory method simulations.

\subsection*{\sffamily \large ACKNOWLEDGMENTS}

This work is supported by the Austrian Science Fund (FWF) through the project P25827.
The authors are specially thankful to the past and present members of the development team of \textsc{Sharc} -- M.~Richter, J.~Gonz\'alez-V\'azquez, I.~Sola, M.~Oppel and M.~Ruckenbauer -- for their many contributions and discussions, and also to the many colleagues who through their questions and suggestions have contributed to improve \textsc{Sharc} from the first to the current implementation.
We are also indebted to T.~M\"uller, T.~Hoffmann-Ostenhof, M.~Persico and G.~Granucci for inspiring discussions. The COST Actions CM1204 (XLIC) and CM1305 (ECOSTBio) are thanked for promoting different collaborative applications related to \textsc{Sharc}.
Finally, the Vienna Scientific Cluster (VSC) is gratefully acknowledged for generous allocation of computer time.

\appendix

\subsection*{Appendix A. Propagation using substeps}

Equation~\eqref{eq:three-step-prop-brief} contains the propagation matrix $\mathbf{P}^\text{MCH}(t+\Delta t,t)$ in the MCH basis.
For small $\Delta t$, this matrix can be calculated from equation~\eqref{eq:three-step-prop}.
For too large $\Delta t$, it is possible to split the interval into $n$ subintervals $\Delta\tau=\frac{\Delta t}{n}$.
Here, we give the corresponding equations for the calculation of $\mathbf{P}^\text{MCH}(t+\Delta t,t)$ using subintervals.

The propagator can be calculated by linearly interpolating $\mathbf{H}^{\text{MCH}}$ and $\mathbf{T}^{\text{MCH}}$, using
\begin{align}
  \mathbf{P}^{\text{MCH}}(t+\Delta t,t)=&
  \prod\limits_{i=1}^{n}
  \mathbf{P}_i,\\
  \mathbf{P}_i=&
  \exp\left[
      -\left(
        \text{i}\mathbf{H}_i
        +\mathbf{T}_i
      \right)\Delta\tau
  \right],\\
  \mathbf{H}_i=&
  \mathbf{H}^{\text{MCH}}(t) + \frac{i}{n}
  \left(
    \mathbf{H}^{\text{MCH}}(t+\Delta t)-\mathbf{H}^{\text{MCH}}(t)
  \right)\label{eq:ham_propn},\\
  \mathbf{T}_i=&
  \mathbf{T}^{\text{MCH}}(t) + \frac{i}{n}
  \left(
    \mathbf{T}^{\text{MCH}}(t+\Delta t)-\mathbf{T}^{\text{MCH}}(t)
  \right),
\end{align}
where the matrix elements of $\mathbf{T}^{\text{MCH}}$ are
\begin{equation}
  \left(\mathbf{T}^{\text{MCH}}(t)\right)_{\beta\alpha}
  =
  \frac{\mathrm{d} \mathbf{R}}{\mathrm{d} t}\cdot
  \left\langle
    \psi_\beta(t)
  \middle|
    \nabla_{\mathbf{R}}
  \middle|
    \psi_\alpha(t)
  \right\rangle
\end{equation}
(classical Tully scheme\cite{Tully1990JCP}) or
\begin{equation}
  \left(\mathbf{T}^{\text{MCH}}(t)\right)_{\beta\alpha}
  =
  \left\langle
    \psi_\beta(t)
  \middle|
    \frac{\mathrm{d}}{\mathrm{d} t}
  \middle|
    \psi_\alpha(t)
  \right\rangle
\end{equation}
(Hammes-Schiffer-Tully scheme\cite{Hammes-Schiffer1994JCP}).

Using the local diabatization scheme,\cite{Granucci2001JCP, Plasser2012JCP} the propagator is, instead, given by
\begin{align}
  \mathbf{P}^{\text{MCH}}(t+\Delta t,t)=&
  \mathbf{S}^{\text{MCH}}(t,t+\Delta t)^\dagger\prod\limits_{i=1}^{n}
  \mathbf{P}_i,\\
  \mathbf{P}_i=&
  \exp\left[
      -\text{i}\mathbf{H}_i\Delta\tau
  \right],\\
  \mathbf{H}_i=&
  \mathbf{H}^{\text{MCH}}(t) + \frac{i}{n}
  \left(
    \mathbf{H}^{\text{MCH}}_{\text{tra}}
    -\mathbf{H}^{\text{MCH}}(t)
  \right)\label{eq:ham_propl},\\
  \mathbf{H}^{\text{MCH}}_{\text{tra}}=&
    \mathbf{S}^{\text{MCH}}(t,t+\Delta t)
    \mathbf{H}^{\text{MCH}}(t+\Delta t)
    \mathbf{S}^{\text{MCH}}(t,t+\Delta t)^\dagger,
\end{align}
where $\mathbf{H}^{\text{MCH}}_{\text{tra}}$ is the matrix representation of $\hat{H}_\text{el}(t+\Delta t)$ in the basis of the states at time $t$.
The overlap matrix $\mathbf{S}^{\text{MCH}}(t,t+\Delta t)$ has elements
\begin{equation}
  \left(\mathbf{S}^{\text{MCH}}(t,t+\Delta t)\right)_{\beta\alpha}=
  \left\langle
    \psi_\beta(t)
  \middle|
    \psi_\alpha(t+\Delta t)
  \right\rangle.
\end{equation}

\subsection*{Appendix B. Continuous transformation-matrix phase}

A Hermitian matrix $\mathbf{H}^{\text{MCH}}$ can always be diagonalized.
Its eigenvectors form the rows of a unitary matrix $\mathbf{U}$, which can be used to transform between the original basis and the basis of the eigenfunctions of $\mathbf{H}$:
\begin{equation}
  \mathbf{H}^{\text{diag}}=\mathbf{U}^\dagger\mathbf{H}^{\text{MCH}}\mathbf{U}.
\end{equation}

However, the condition that $\mathbf{U}$ diagonalizes $\mathbf{H}^{\text{MCH}}$ is not sufficient to define $\mathbf{U}$ uniquely. Each normalized eigenvector can be multiplied by a complex number on the unit circle and still remains a normalized eigenvector.
We define a second transformation matrix $\tilde{\mathbf{U}}=\mathbf{U}\boldsymbol{\Phi}$, where $\boldsymbol{\Phi}$ is a diagonal matrix with elements $\Phi_{\beta\alpha}=\delta_{\beta\alpha}\text{e}^{\text{i}\phi_\alpha}$ (i.e., an eigenvector of $\tilde{\mathbf{U}}$ is the same eigenvector of $\mathbf{U}$ times a complex phase).
By this definition, $\boldsymbol{\Phi}$ is unitary ($\boldsymbol{\Phi}^\dagger\boldsymbol{\Phi}=\mathbf{1}$) and commutes with $\mathbf{H}^\text{diag}$, since two diagonal matrices always commute.
With this properties it can be shown by
\begin{equation}
  \tilde{\mathbf{U}}^\dagger\mathbf{H}^{\text{MCH}}\tilde{\mathbf{U}}
  =
  \boldsymbol{\Phi}^\dagger\mathbf{U}^\dagger\mathbf{H}^\text{MCH}\mathbf{U}\boldsymbol{\Phi}
  =
  \boldsymbol{\Phi}^\dagger\mathbf{H}^\text{diag}\boldsymbol{\Phi}
  =
  \boldsymbol{\Phi}^\dagger\boldsymbol{\Phi}\mathbf{H}^\text{diag}
  =
  \mathbf{H}^\text{diag}
\end{equation}
that also $\tilde{\mathbf{U}}$ diagonalizes $\mathbf{H}^{\text{MCH}}$ and produces the same eigenvalues.

In most libraries for numerical diagonalization of Hermitian matrices, the eigenvector phases given in $\boldsymbol{\Phi}$ are either uncontrolled (so that the phases depend on the details of the implementation) or are fixed by some constraints (e.g., the first non-zero element of the eigenvector is made real-valued).
However, by these conditions the eigenvector phases are not continuous in time -- a small change in geometry from $\mathbf{R}(t)$ to $\mathbf{R}(t+\Delta t)$ may lead to a large jump in some eigenvector phases.
Since the transformation matrix with discontinuous phases $\mathbf{U}(t)\boldsymbol{\Phi}(t)$ is itself discontinuous, it is not differentiable.
Thus, it is not possible to calculate the antihermitian term $\mathbf{U}^\dagger\frac{\mathrm{d}}{\mathrm{d} t}\mathbf{U}$ from equation~\eqref{eq:eom_diag}, neither analytically nor by finite differences.
Attempting to calculate finite differences anyways will lead to a result that is not antihermitian and generally overestimates the size of the matrix elements of the differential.
Thus, attempting to use equation~\eqref{eq:eom_diag} for the propagation of the wavefunction and relying on the finite-difference calculation of $\mathbf{U}^\dagger\frac{\mathrm{d}}{\mathrm{d} t}\mathbf{U}$ without making the eigenvector phases (approximately) continuous will lead both to overestimated population transfer -- possibly at any time -- and to violation of the conservation of the wavefunction norm.

In order to facilitate the calculation of $\mathbf{U}^\dagger\frac{\mathrm{d}}{\mathrm{d} t}\mathbf{U}$, it is necessary to control the phase of the transformation matrix $\mathbf{U}$.
Specifically, this means that the phases of $\mathbf{U}(t+\Delta t)$ have to be aligned to the phases of $\mathbf{U}(t)$.
Hence, given $\mathbf{U}(t)$ and $\tilde{\mathbf{U}}(t+\Delta t)$, we need to find $\boldsymbol{\Phi}$, which allows to obtain the phase-aligned transformation matrix $\mathbf{U}(t+\Delta t)$:
\begin{equation}
  \mathbf{U}(t+\Delta t)
  =
  \tilde{\mathbf{U}}(t+\Delta t)\boldsymbol{\Phi}^\dagger.
\end{equation}
Choosing $\boldsymbol{\Phi}^\dagger=\tilde{\mathbf{U}}^\dagger(t+\Delta t)\mathbf{U}(t)$ would trivially accomplish the goal of phase alignment, since then $\mathbf{U}(t+\Delta t)=\mathbf{U}(t)$.
However, this $\boldsymbol{\Phi}^\dagger$ does not commute with $\mathbf{H}^\text{diag}(t+\Delta t)$.
Hence, we propose the following ansatz for $\boldsymbol{\Phi}^\dagger$:
\begin{equation}
  \boldsymbol{\Phi}^\dagger
  =
  \hat{\mathcal{O}}
  \hat{\mathcal{C}}
  \left[
    \tilde{\mathbf{U}}^\dagger(t+\Delta t)
    \mathbf{U}(t)
  \right]
  =
  \hat{\mathcal{O}}
  \hat{\mathcal{C}}
  \mathbf{V}.
  \label{eq:continuation}
\end{equation}
Here, the operator $\hat{\mathcal{O}}$ ensures that $\boldsymbol{\Phi}^\dagger$ is orthogonal/unitary ($\boldsymbol{\Phi}^\dagger\boldsymbol{\Phi}=\mathbf{1}$) and $\hat{\mathcal{C}}$ ensures that $\boldsymbol{\Phi}^\dagger$ commutes with $\mathbf{H}^\text{diag}$ ($\mathbf{H}^\text{diag}\boldsymbol{\Phi}^\dagger=\boldsymbol{\Phi}^\dagger\mathbf{H}^\text{diag}$).

The commutation property can be achieved by defining
\begin{equation}
  \hat{\mathcal{C}}\mathbf{V}
  =
  \sum\limits_{\mathcal{S}_i} \hat{\mathcal{P}}(\mathcal{S}_i)\mathbf{V}\hat{\mathcal{P}}(\mathcal{S}_i),
  \label{eq:projector}
\end{equation}
where $\hat{\mathcal{P}}(\mathcal{S}_i)$ is the projection operator on the $i$-th eigenspace $\mathcal{S}_i$ of $\mathbf{H}^\text{diag}(t+\Delta t)$:
\begin{equation}
  \hat{\mathcal{P}}(\mathcal{S}_i)
  =
  \sum\limits_{\mathbf{j}\in\mathcal{S}_i} \mathbf{j}\mathbf{j}^\dagger.
\end{equation}
Here, the vectors $\mathbf{j}$ are the eigenvectors of $\mathbf{H}^\text{diag}(t+\Delta t)$.
Because $\mathbf{H}^\text{diag}$ is by definition diagonal, the elements of $\hat{\mathcal{C}}\mathbf{V}$ can be written very compactly as
\begin{equation}
  \left(\hat{\mathcal{C}}\mathbf{V}\right)_{\beta\alpha}
  =
  V_{\beta\alpha}
  \delta(H_{\beta\beta}^\text{diag}-H_{\alpha\alpha}^\text{diag}).
  \label{eq:projector2}
\end{equation}
It is easy to show that with this definition $\hat{\mathcal{C}}\mathbf{V}\mathbf{H}^\text{diag}=\mathbf{H}^\text{diag}\hat{\mathcal{C}}\mathbf{V}$.

The property $\boldsymbol{\Phi}^\dagger\boldsymbol{\Phi}=\mathbf{1}$ can be achieved if the operator $\hat{\mathcal{O}}$ orthonormalizes $\hat{\mathcal{C}}\mathbf{V}$.
The optimal orthonormalization scheme here is L\"owdin's symmetric orthonormalization,\cite{Lowdin1950JCP} since it changes $\hat{\mathcal{C}}\mathbf{V}$ as little as possible.\cite{Carlson1957PR}

The above-described algorithm leads for $\Delta t\rightarrow0$ to $\mathbf{U}(t+\Delta t)\rightarrow\mathbf{U}(t)$.
However, for finite $\Delta t$, one has to take some care.
For the special case that the eigenvalues are non-degenerate ($H_{\alpha\alpha}^\text{diag}\neq H_{\beta\beta}^\text{diag}$ for all $\alpha, \beta$), one can show that
\begin{equation}
  U_{\beta\alpha}(t+\Delta t)
  =
  \sum\limits_i
  \tilde{U}_{\beta i}(t+\Delta t)
  \frac{
    \sum\limits_j
    \tilde{U}^*_{ji}(t+\Delta t)
    U_{j\alpha}(t)
    \delta_{i\alpha}
  }{
    \left|
      \sum\limits_j
      \tilde{U}^*_{ji}(t+\Delta t)
      U_{j\alpha}(t)
      \delta_{i\alpha}
    \right|
  }.
\end{equation}
Using $\tilde{U}_{\beta\alpha}(t+\Delta t)=U_{\beta\alpha}(t+\Delta t)\text{e}^{\text{i}\phi_\alpha}$ and reducing the $\delta_{i\alpha}$ gives:
\begin{equation}
  U_{\beta\alpha}(t+\Delta t)
  =
  U_{\beta\alpha}(t+\Delta t)
  \frac{
    \sum\limits_j
    U^*_{j\alpha}(t+\Delta t)
    U_{j\alpha}(t)
  }{
    \left|
      \sum\limits_j
      U^*_{j\alpha}(t+\Delta t)
      U_{j\alpha}(t)
    \right|
  }
  =
  U_{\beta\alpha}(t+\Delta t)
  \frac{\mathbf{u}_\alpha(t+\Delta t)\cdot\mathbf{u}_\alpha(t)}{|\mathbf{u}_\alpha(t+\Delta t)\cdot\mathbf{u}_\alpha(t)|},
\end{equation}
where $\mathbf{u}_\alpha$ is the $\alpha$-th column vector of $\mathbf{U}$.
It is important to stress here that this only holds if $\mathbf{u}_\alpha(t+\Delta t)\cdot\mathbf{u}_\alpha(t)$ is real-valued, which will in general only be approximately true if no root flipping occurred between $t$ and $t+\Delta t$ (i.e., the $\alpha$-th eigenvector at time $t+\Delta t$ is the same as the $\alpha$-th eigenvector at time $t$).
Thus, for finite timesteps $\Delta t$ we have to require that any root flipping is removed from $\mathbf{V}=\tilde{\mathbf{U}}^\dagger(t+\Delta t)\mathbf{U}(t)$.
This can be accomplished by a simple algorithm that takes the matrix $\mathbf{V}$ and brings it into diagonally dominant form by reordering its columns.

Finally, it is crucial for the algorithm that $\Delta t$ is small.
A criterion for smallness can be found from the antihermicity property of $\mathbf{U}^\dagger\frac{\mathrm{d}}{\mathrm{d} t}\mathbf{U}$.
This term can be calculated by finite differences as
\begin{equation}
  \mathbf{U}^\dagger\frac{\mathrm{d}}{\mathrm{d} t}\mathbf{U}
  \approx
  \mathbf{U}^\dagger(t+\Delta t)
  \frac{\mathbf{U}(t+\Delta t)-\mathbf{U}(t)}{\Delta t}
  =\frac{1}{\Delta t}\left(\mathbf{1}-\mathbf{U}^\dagger(t+\Delta t)\mathbf{U}(t)\right).
\end{equation}
The antihermicity of $\mathbf{U}^\dagger\frac{\mathrm{d}}{\mathrm{d} t}\mathbf{U}$ requires that its diagonal elements be zero, and hence it can be followed that the diagonal elements of $\mathbf{U}^\dagger(t+\Delta t)\mathbf{U}(t)$ need to be close to 1.
Hence, we can choose a threshold $\varepsilon$ and require that
\begin{equation}
  1-\varepsilon
  <
  \left(\mathbf{U}^\dagger(t+\Delta t)\mathbf{U}(t)\right)_{\alpha\alpha} \forall \alpha.
  \label{eq:dt_condition}
\end{equation}
If the condition is not fulfilled, a smaller value of $\Delta t$ is required.

The complete algorithm can thus be given by the following steps:
\begin{enumerate}
  \item Numerically diagonalize: $\mathbf{H}^{\text{diag}}(t+\Delta t)=\tilde{\mathbf{U}}^\dagger(t+\Delta t)\mathbf{H}^{\text{MCH}}(t+\Delta t)\tilde{\mathbf{U}}(t+\Delta t)$.
  \item Calculate $\mathbf{V}=\tilde{\mathbf{U}}^\dagger(t+\Delta t)\mathbf{U}(t)$.
  \item Bring $\mathbf{V}$ into diagonally dominant form by exchanging columns.
  \item Obtain the matrix $\hat{\mathcal{C}}\mathbf{V}$ from equation~\eqref{eq:projector2}.
  \item Orthonormalize $\hat{\mathcal{C}}\mathbf{V}$ by L\"owdin's symmetric orthonormalization.\cite{Lowdin1950JCP} The result is $\boldsymbol{\Phi}^\dagger$.
  \item Obtain the phase-controlled matrix $\mathbf{U}(t+\Delta t)=\tilde{\mathbf{U}}(t+\Delta t)\boldsymbol{\Phi}^\dagger$.
  \item Check whether condition~\eqref{eq:dt_condition} holds. If not, divide the interval $[t,t+\Delta t]$ into subintervals, linearly interpolate $\mathbf{H}^{\text{MCH}}$ and carry out steps~1 to 6 for each subinterval. If the subintervals would be shorter than a given threshold, do not subdivide (termination of the recursive algorithm).
\end{enumerate}

Finally, it should be stated that the algorithm outlined above also considers that the Hamiltonian can have degenerate eigenvalues.
In this case, the mixing angles between the degenerate states are also not uniquely defined, in addition to the eigenvector phases.
In step~4, the operator $\hat{\mathcal{C}}$ (see equation~\eqref{eq:projector2}) keeps off-diagonal elements of $\mathbf{V}$ corresponding to degenerate eigenfunctions, which leads to the removal of the arbitrariness in the mixing angles.


\begin{thebibliography}{100}

\bibitem{Klessinger1995}
M.~Klessinger, J.~Michl: \emph{Excited States and Photochemistry of Organic
  Molecules}. VCH Publishers Inc., New York (1995).

\bibitem{McQuarrie1997}
D.~A. McQuarrie, J.~D. Simon: \emph{Physical Chemistry}. University Science
  Books (1997).

\bibitem{Turro2009}
N.~J. Turro, J.~C. Scaiano, V.~Ramamurthy: \emph{Principles of Molecular
  Photochemistry: An Introduction}. University Science Books, Sausalito, CA
  (2009).

\bibitem{Kasha1950DFS}
M.~Kasha: \emph{Discuss. Faraday Soc.}, \textbf{9}, 14 (1950).

\bibitem{Domcke2004}
W.~Domcke, D.~R. Yarkony, H.~K\"oppel (editors): \emph{Conical Intersections:
  Electronic Structure, Dynamics and Spectroscopy}. World Scientific Publishing
  (2004).

\bibitem{Schoenlein1991S}
R.~Schoenlein, L.~Peteanu, R.~Mathies, C.~Shank: \emph{Science}, \textbf{254},
  412 (1991).

\bibitem{Frutos2007PNAS}
L.~M. Frutos, T.~Andruni\'ow, F.~Santoro, N.~Ferr\'e, M.~Olivucci: \emph{Proc.
  Natl. Acad. Sci.}, \textbf{104}, 7764 (2007).

\bibitem{Crespo-Hernandez2004CR}
C.~E. Crespo-Hern\'{a}ndez, B.~Cohen, P.~M. Hare, B.~Kohler: \emph{Chem. Rev.},
  \textbf{104}, 1977 (2004).

\bibitem{Barbatti2014TCC}
M.~Barbatti, A.~C. Borin, S.~Ullrich: In \emph{Photoinduced Phenomena in
  Nucleic Acids I}, volume 355 of \emph{Topics in Current Chemistry}, Springer
  Berlin Heidelberg, 1--32 (2014).

\bibitem{Weinkauf2002EPJD}
R.~Weinkauf, J.-P. Schermann, M.~de~Vries, K.~Kleinermanns: \emph{Eur. Phys. J.
  D}, \textbf{20}, 309 (2002).

\bibitem{Stephansen2012JACS}
A.~B. Stephansen, R.~Y. Brogaard, T.~S. Kuhlman, L.~B. Klein, J.~B.
  Christensen, T.~I. S{\o}lling: \emph{J. Am. Chem. Soc.}, \textbf{134}, 20279
  (2012).

\bibitem{Perot2010JPCA}
M.~P\'erot, B.~Lucas, M.~Barat, J.~A. Fayeton, C.~Jouvet: \emph{J. Phys. Chem.
  A}, \textbf{114}, 3147 (2010).

\bibitem{Vries2007ARPC}
M.~S. de~Vries, P.~Hobza: \emph{Annu. Rev. Phys. Chem.}, \textbf{58}, 585
  (2007).

\bibitem{Seinfeld1997}
J.~H. Seinfeld, S.~N. Pandis: \emph{Atmospheric Chemistry and Physics from Air
  Pollution to Climate Change}. John Wiley \& Sons, Inc. (1997).

\bibitem{Heicklen1980RCI}
J.~Heicklen, N.~Kelly, K.~Partymiller: \emph{Res. Chem. Intermed.}, \textbf{3},
  315 (1980).

\bibitem{Wilkinson2010ARPCSC}
I.~Wilkinson, B.~J.~Whitaker: \emph{Annu. Rep. Prog. Chem., Sect. C},
  \textbf{106}, 274 (2010).

\bibitem{Davis1993JPC}
H.~F. Davis, B.~Kim, H.~S. Johnston, Y.~T. Lee: \emph{J. Phys. Chem.},
  \textbf{97}, 2172 (1993).

\bibitem{Gutlich1994ACIEE}
P.~G\"utlich, A.~Hauser, H.~Spiering: \emph{Angew. Chem., Int. Ed. Engl.},
  \textbf{33}, 2024 (1994).

\bibitem{Roy2000APL}
S.~Roy, C.~P. Singh, K.~P.~J. Reddy: \emph{Appl. Phys. Lett.}, \textbf{77},
  2656 (2000).

\bibitem{Epstein2003MB}
A.~J. Epstein: \emph{MRS Bulletin}, \textbf{28}, 492 (2003).

\bibitem{Ito2014ACIE}
A.~Ito, A.~Shimizu, N.~Kishida, Y.~Kawanaka, D.~Kosumi, H.~Hashimoto, Y.~Teki:
  \emph{Angew. Chem. Int. Ed.}, \textbf{53}, 6715 (2014).

\bibitem{Segura2005CSR}
J.~L. Segura, N.~Martin, D.~M. Guldi: \emph{Chem. Soc. Rev.}, \textbf{34}, 31
  (2005).

\bibitem{Uoyama2012N}
H.~Uoyama, K.~Goushi, K.~Shizu, H.~Nomura, C.~Adachi: \emph{Nature},
  \textbf{492}, 234 (2012).

\bibitem{Hsu2012JACS}
C.-C. Hsu, C.-C. Lin, P.-T. Chou, C.-H. Lai, C.-W. Hsu, C.-H. Lin, Y.~Chi:
  \emph{J. Am. Chem. Soc.}, \textbf{134}, 7715 (2012).

\bibitem{Gust1993JACS}
D.~Gust, T.~A. Moore, A.~L. Moore, A.~A. Krasnovsky, P.~A. Liddell, D.~Nicodem,
  J.~M. DeGraziano, P.~Kerrigan, L.~R. Makings, P.~J. Pessiki: \emph{J. Am.
  Chem. Soc.}, \textbf{115}, 5684 (1993).

\bibitem{Saito2014JPCB}
T.~Saito, W.~Thiel: \emph{J. Phys. Chem. B}, \textbf{118}, 5034 (2014).

\bibitem{Dolmans2003NRC}
D.~E. J. G.~J. Dolmans, D.~Fukumura, R.~K. Jain: \emph{Nat. Rev. Cancer},
  \textbf{3}, 380 (2003).

\bibitem{Griesbeck2004ACR}
A.~G. Griesbeck, M.~Abe, S.~Bondock: \emph{Acc. Chem. Res.}, \textbf{37}, 919
  (2004).

\bibitem{Zapata-Rivera2014CEJ}
J.~Zapata-Rivera, R.~Caballol, C.~J. Calzado, D.~G. Liakos, F.~Neese:
  \emph{Chem. Eur. J.}, \textbf{20}, 13296 (2014).

\bibitem{Roundhill1994}
D.~M. Roundhill: \emph{Photochemistry and Photophysics of Metal Complexes}.
  Springer US (1994).

\bibitem{Chergui2012DT}
M.~Chergui: \emph{Dalton Trans.}, \textbf{41}, 13022 (2012).

\bibitem{Gawelda2007JACS}
W.~Gawelda, A.~Cannizzo, V.-T. Pham, F.~van Mourik, C.~Bressler, M.~Chergui:
  \emph{J. Am. Chem. Soc.}, \textbf{129}, 8199 (2007).

\bibitem{Yersin2004TCC}
H.~Yersin: \emph{Top. Curr. Chem.}, \textbf{241}, 1 (2004).

\bibitem{Cannizzo2006ACIE}
A.~Cannizzo, F.~van Mourik, W.~Gawelda, G.~Zgrablic, C.~Bressler, M.~Chergui:
  \emph{Angew. Chem. Int. Ed.}, \textbf{45}, 3174 (2006).

\bibitem{Cannizzo2008JACSa}
A.~Cannizzo, A.~M. Blanco-Rodr\'{\i}guez, A.~El~Nahhas, J.~\v{S}ebera,
  S.~Z\'ali\v{a}, A.~Vl\v{c}ek, Jr., M.~Chergui: \emph{J. Am. Chem. Soc.},
  \textbf{130}, 8967 (2008).

\bibitem{Bram2013JPCC}
O.~Br\"am, F.~Messina, E.~Baranoff, A.~Cannizzo, M.~K. Nazeeruddin, M.~Chergui:
  \emph{J. Phys. Chem. C}, \textbf{117}, 15958 (2013).

\bibitem{Amaral2010JCP}
G.~A. Amaral, A.~Arregui, L.~Rubio-Lago, J.~D. Rodr\'{\i}guez, L.~Ba\~nares:
  \emph{J. Chem. Phys.}, \textbf{133}, 064303 (2010).

\bibitem{Parker2009CPL}
D.~S.~N. Parker, R.~S. Minns, T.~J. Penfold, G.~A. Worth, H.~H. Fielding:
  \emph{Chem. Phys. Lett.}, \textbf{469}, 43 (2009).

\bibitem{Cavaleri1996CPL}
J.~J. Cavaleri, K.~Prater, R.~M. Bowman: \emph{Chem. Phys. Lett.},
  \textbf{259}, 495 (1996).

\bibitem{Ohshima2003JPCA}
Y.~Ohshima, T.~Fujii, T.~Fujita, D.~Inaba, M.~Baba: \emph{J. Phys. Chem. A},
  \textbf{107}, 8851 (2003).

\bibitem{Satzger2004JPCA}
H.~Satzger, B.~Schmidt, C.~Root, W.~Zinth, B.~Fierz, F.~Krieger, T.~Kiefhaber,
  P.~Gilch: \emph{J. Phys. Chem. A}, \textbf{108}, 10072 (2004).

\bibitem{Ramesdonk2006JPCA}
H.~J. van Ramesdonk, B.~H. Bakker, M.~M. Groeneveld, J.~W. Verhoeven, B.~D.
  Allen, J.~P. Rostron, A.~Harriman: \emph{J. Phys. Chem. A}, \textbf{110},
  13145 (2006).

\bibitem{Reichardt2009JCP}
C.~Reichardt, R.~A. Vogt, C.~E. Crespo-Hern{\'{a}}ndez: \emph{J. Chem. Phys.},
  \textbf{131}, 224518 (2009).

\bibitem{Yang2012JCP}
C.~Yang, H.~Su, X.~Sun, M.~W. George: \emph{J. Chem. Phys.}, \textbf{136},
  204507 (2012).

\bibitem{Tamai1992CPL}
N.~Tamai, T.~Asahi, H.~Masuhara: \emph{Chem. Phys. Lett.}, \textbf{198}, 413
  (1992).

\bibitem{Wolf2012M}
T.~J.~A. Wolf, D.~Voll, C.~Barner-Kowollik, A.-N. Unterreiner:
  \emph{Macromolecules}, \textbf{45}, 2257 (2012).

\bibitem{Ma2007CEJ}
C.~Ma, Y.~Du, W.~M. Kwok, D.~L. Phillips: \emph{Chem. Eur. J.}, \textbf{13},
  2290 (2007).

\bibitem{Park2006JCP}
S.~T. Park, J.~S. Feenstra, A.~H. Zewail: \emph{J. Chem. Phys.}, \textbf{124},
  174707 (2006).

\bibitem{Ou2013JPCC}
Q.~Ou, J.~E. Subotnik: \emph{J. Phys. Chem. C}, \textbf{117}, 19839 (2013).

\bibitem{Schalk2014JPCA}
O.~Schalk, M.~S. Schuurman, G.~Wu, P.~Lang, M.~Mucke, R.~Feifel, A.~Stolow:
  \emph{J. Phys. Chem. A}, \textbf{118}, 2279 (2014).

\bibitem{Huix-Rotllant2013PCCP}
M.~Huix-Rotllant, D.~Siri, N.~Ferr{\'{e}}: \emph{Phys. Chem. Chem. Phys.},
  \textbf{15}, 19293 (2013).

\bibitem{Vogt2013JPCA}
R.~A. Vogt, C.~Reichardt, C.~E. Crespo-Hern{\'{a}}ndez: \emph{J. Phys. Chem.
  A}, \textbf{117}, 6580 (2013).

\bibitem{Morales-Cueto2007JPCA}
R.~Morales-Cueto, M.~Esquivelzeta-Rabell, J.~S. Zugazagoitia, J.~Peon: \emph{J.
  Phys. Chem. A}, \textbf{111}, 552 (2007).

\bibitem{Zugazagoitia2008JPCA}
J.~S. Zugazagoitia, C.~X. {Almora-D\'iaz}, J.~Peon: \emph{J. Phys. Chem. A},
  \textbf{112}, 358 (2008).

\bibitem{Zugazagoitia2009JPCA}
J.~S. Zugazagoitia, E.~Collado-Fregoso, E.~F. Plaza-Medina, J.~Peon: \emph{J.
  Phys. Chem. A}, \textbf{113}, 805 (2009).

\bibitem{Lopez-Arteaga2013JPCB}
R.~L{\'{o}}pez-Arteaga, A.~B. Stephansen, C.~A. Guarin, T.~I. S{\o}lling,
  J.~Peon: \emph{J. Phys. Chem. B}, \textbf{117}, 9947 (2013).

\bibitem{Plaza-Medina2011JPCA}
E.~F. Plaza-Medina, W.~Rodr{\'{i}}guez-C{\'{o}}rdoba, J.~Peon: \emph{J. Phys.
  Chem. A}, \textbf{115}, 9782 (2011).

\bibitem{Mohammed2008JPCA}
O.~F. Mohammed, E.~Vauthey: \emph{J. Phys. Chem. A}, \textbf{112}, 3823 (2008).

\bibitem{Collodo-Fregoso2009JPCA}
E.~Collodo-Fregoso, J.~S. Zugazagoitia, E.~F. Plaza-Medina, J.~Peon: \emph{J.
  Phys. Chem. A}, \textbf{113}, 13498 (2009).

\bibitem{Ghosh2012JPCA}
R.~Ghosh, D.~K. Palit: \emph{J. Phys. Chem. A}, \textbf{116}, 1993 (2012).

\bibitem{Crespo-Hernandez2008JPCA}
C.~E. Crespo-Hern\'andez, G.~Burdzinski, R.~Arce: \emph{J. Phys. Chem. A},
  \textbf{112}, 6313 (2008).

\bibitem{Reichardt2011JPCB}
C.~Reichardt, C.~Guo, C.~E. Crespo-Hern\'andez: \emph{J. Phys. Chem. B},
  \textbf{115}, 3263 (2011).

\bibitem{Reichardt2010JPCL}
C.~Reichardt, C.~E. Crespo-Hern\'{a}ndez: \emph{J. Phys. Chem. Lett.},
  \textbf{1}, 2239 (2010).

\bibitem{Reichardt2010CC}
C.~Reichardt, C.~E. Crespo-Hern\'{a}ndez: \emph{Chem. Commun.}, \textbf{46},
  5963 (2010).

\bibitem{Pollum2014JCP}
M.~Pollum, C.~E. Crespo-Hern\'andez: \emph{J. Chem. Phys.}, \textbf{140},
  071101 (2014).

\bibitem{Martinez-Fernandez2012CC}
L.~Mart{\'i}nez-Fern{\'a}ndez, L.~Gonz{\'a}lez, I.~Corral: \emph{Chem.
  Commun.}, \textbf{48}, 2134 (2012).

\bibitem{Cui2014JPCL}
G.~Cui, W.~Thiel: \emph{J. Phys. Chem. Lett.}, \textbf{5}, 2682 (2014).

\bibitem{Gobbo2014CTC}
J.~P. Gobbo, A.~C. Borin: \emph{Comput. Theor. Chem.}, \textbf{1040-1041}, 195
  (2014).

\bibitem{Kobayashi2009JPCA}
T.~Kobayashi, H.~Kuramochi, Y.~Harada, T.~Suzuki, T.~Ichimura: \emph{J. Phys.
  Chem. A}, \textbf{113}, 12088 (2009).

\bibitem{Dietz1987JACS}
T.~M. Dietz, R.~J. Von~Trebra, B.~J. Swanson, T.~H. Koch: \emph{J. Am. Chem.
  Soc.}, \textbf{109}, 1793 (1987).

\bibitem{Cadet1990}
J.~Cadet, P.~Vigny: In H.~Morrison (editor), \emph{Bioorganic Photochemistry 1:
  Photochemistry and the Nucleic Acids}, Wiley-Interscience (1990).

\bibitem{Schreier2007S}
W.~J. Schreier, T.~E. Schrader, F.~O. Koller, P.~Gilch, C.~E.
  Crespo-Hern\'{a}ndez, V.~N. Swaminathan, T.~Charell, W.~Zinth, B.~Kohler:
  \emph{Science}, \textbf{315}, 625 (2007).

\bibitem{Barbatti2014CPC}
M.~Barbatti: \emph{ChemPhysChem}, \textbf{15}, 3342 (2014).

\bibitem{Marian2012WCMS}
C.~M. Marian: \emph{WIREs Comput. Mol. Sci.}, \textbf{2}, 187 (2012).

\bibitem{Grimme1998CPL}
S.~Grimme, M.~Woeller, S.~D. Peyerimhoff, D.~Danovich, S.~Shaik: \emph{Chem.
  Phys. Lett.}, \textbf{287}, 601  (1998).

\bibitem{Harvey2000CPL}
J.~N. Harvey, S.~Grimme, M.~Woeller, S.~D. Peyerimhoff, D.~Danovich, S.~Shaik:
  \emph{Chem. Phys. Lett.}, \textbf{322}, 358  (2000).

\bibitem{Tatchen2007PCCP}
J.~Tatchen, N.~Gilka, C.~M. Marian: \emph{Phys. Chem. Chem. Phys.}, \textbf{9},
  5209 (2007).

\bibitem{Cui1999JCP}
Q.~Cui, K.~Morokuma, J.~M. Bowman, S.~J. Klippenstein: \emph{J. Chem. Phys.},
  \textbf{110}, 9469 (1999).

\bibitem{Veldman2008JPCA}
D.~Veldman, S.~M.~A. Chopin, S.~C.~J. Meskers, R.~A.~J. Janssen: \emph{J. Phys.
  Chem. A}, \textbf{112}, 8617 (2008).

\bibitem{Bargheer2002PCCP}
M.~Bargheer, M.~G{\"u}hr, P.~Dietrich, N.~Schwentner: \emph{Phys. Chem. Chem.
  Phys.}, \textbf{4}, 75 (2002).

\bibitem{Korolkov2004JCP}
M.~V. Korolkov, J.~Manz: \emph{J. Chem. Phys.}, \textbf{120}, 11522 (2004).

\bibitem{Korolkov2004CPL}
M.~V. Korolkov, J.~Manz: \emph{Chem. Phys. Lett.}, \textbf{393}, 44  (2004).

\bibitem{Daniel1995JCP}
C.~Daniel, M.-C. Heitz, J.~Manz, C.~Ribbing: \emph{J. Chem. Phys.},
  \textbf{102}, 905 (1995).

\bibitem{Heitz1995CCR}
M.-C. Heitz, K.~Finger, C.~Daniel: \emph{Coord. Chem. Rev.}, \textbf{159}, 171
  (1995).

\bibitem{Daniel1996IJQC}
C.~Daniel, R.~de~Vivie-Riedle, M.-C. Heitz, J.~Manz, P.~Saalfrank: \emph{Int.
  J. Quant. Chem.}, \textbf{57}, 595 (1996).

\bibitem{Daniel1999JPCA}
C.~Daniel, D.~Guillaumont, C.~Ribbing, B.~Minaev: \emph{J. Phys. Chem. A},
  \textbf{103}, 5766 (1999).

\bibitem{Bruand-Cote2002CEJ}
I.~Bruand-Cote, C.~Daniel: \emph{Chem. Eur. J.}, \textbf{8}, 1361 (2002).

\bibitem{Amor2007CP}
N.~B. Amor, D.~Ambrosek, C.~Daniel, R.~Marquardt: \emph{Chem. Phys.},
  \textbf{338}, 81 (2007).

\bibitem{Costa2008NJC}
P.~J. Costa, M.~J. Calhorda, S.~Villaume, C.~Daniel: \emph{New J. Chem.},
  \textbf{32}, 1904 (2008).

\bibitem{Heydova2012JPCA}
R.~Heydov\'{a}, E.~Gindensperger, R.~Romano, J.~S\'{y}kora,
  J.~Anton\'{i}n~Vl\u{c}ek, S.~Za\'{l}is, C.~Daniel: \emph{J. Phys. Chem. A},
  \textbf{116}, 11319 (2012).

\bibitem{Gourlaouen2014DT}
C.~Gourlaouen, C.~Daniel: \emph{Dalton Trans.}, \textbf{43}, 17806 (2014).

\bibitem{Brahim2014CTC}
H.~Brahim, C.~Daniel: \emph{Comp. Theor. Chem.}, \textbf{1040-1041}, 219
  (2014).

\bibitem{Kayanuma2011CCR}
M.~Kayanuma, C.~Daniel, H.~K\"{o}ppel, E.~Gindensperger: \emph{Coord. Chem.
  Rev.}, \textbf{255}, 2693 (2011).

\bibitem{Ando2012CPL}
H.~Ando, S.~Iuchi, H.~Sato: \emph{Chem. Phys. Lett.}, \textbf{535}, 177
  (2012).

\bibitem{Capano2014JPCA}
G.~Capano, M.~Chergui, U.~Rothlisberger, I.~Tavernelli, T.~J. Penfold: \emph{J.
  Phys. Chem. A}, \textbf{118}, 9861 (2014).

\bibitem{Zhao2013JCP}
J.~Zhao: \emph{J. Chem. P}, \textbf{138}, 134309 (2013).

\bibitem{Minns2010PCCP}
R.~S. Minns, D.~S.~N. Parker, T.~J. Penfold, G.~A. Worth, H.~H. Fielding:
  \emph{Phys. Chem. Chem. Phys.}, \textbf{12}, 15607 (2010).

\bibitem{Cohen2007CPL}
A.~Cohen, R.~B. Gerber: \emph{Chem. Phys. Lett.}, \textbf{441}, 48 (2007).

\bibitem{Leveque2014JCP_ISC}
C.~L\'ev\^eque, R.~Ta\"{\i}eb, H.~K\"oppel: \emph{J. Chem. Phys.},
  \textbf{140}, 091101 (2014).

\bibitem{Fu2012JCP}
B.~Fu, Y.-C. Han, J.~M. Bowman, F.~Leonori, N.~Balucani, L.~Angelucci,
  A.~Occhiogrosso, R.~Petrucci, P.~Casavecchia: \emph{J. Chem. Phys.},
  \textbf{137}, 22A532 (2012).

\bibitem{Fu2012PNAS}
B.~Fu, Y.-C. Han, J.~M. Bowman, L.~Angelucci, N.~Balucani, F.~Leonori,
  P.~Casavecchia: \emph{Proc. Natl. Acad. Sci.}, \textbf{109}, 9733 (2012).

\bibitem{Rajak2014JCP}
K.~Rajak, B.~Maiti: \emph{J. Chem. Phys.}, \textbf{140}, 044314 (2014).

\bibitem{Hu2008JPCA}
W.~Hu, G.~Lendvay, B.~Maiti, G.~C. Schatz: \emph{J. Phys. Chem. A},
  \textbf{112}, 2093 (2008).

\bibitem{Tachikawa1997JPCA}
H.~Tachikawa, K.~Ohnishi, T.~Hamabayashi, H.~Yoshida: \emph{J. Phys. Chem. A},
  \textbf{101}, 2229 (1997).

\bibitem{Yamashita1990CPL}
K.~Yamashita, K.~Morokuma: \emph{Chem. Phys. Lett.}, \textbf{169}, 263 (1990).

\bibitem{Maiti2004JPCA}
B.~Maiti, G.~C. Schatz, G.~Lendvay: \emph{J. Phys. Chem. A}, \textbf{108}, 8772
  (2004).

\bibitem{Carbogno2010PRB}
C.~Carbogno, J.~Behler, K.~Reuter, A.~Gro\ss{}: \emph{Phys. Rev. B},
  \textbf{81}, 035410 (2010).

\bibitem{Mai2014JCP_SO2}
S.~Mai, P.~Marquetand, L.~Gonz\'alez: \emph{J. Chem. Phys.}, \textbf{140},
  204302 (2014).

\bibitem{Cui2014JCP}
G.~Cui, W.~Thiel: \emph{J. Chem. Phys.}, \textbf{141}, 124101 (2014).

\bibitem{Favero2013PCCP}
L.~Favero, G.~Granucci, M.~Persico: \emph{Phys. Chem. Chem. Phys.},
  \textbf{15}, 20651 (2013).

\bibitem{Shemesh2013JPCA}
D.~Shemesh, Z.~Lan, R.~B. Gerber: \emph{J. Phys. Chem. A}, \textbf{117}, 11711
  (2013).

\bibitem{Warshel1975CPL}
A.~Warshel, M.~Karplus: \emph{Chem. Phys. Lett.}, \textbf{32}, 11  (1975).

\bibitem{Martinez-Fernandez2014CS}
L.~Mart\'{\i}nez-Fern\'andez, I.~Corral, G.~Granucci, M.~Persico: \emph{Chem.
  Sci.}, \textbf{5}, 1336 (2014).

\bibitem{Richter2012JPCL}
M.~Richter, P.~Marquetand, J.~Gonz\'alez-V\'azquez, I.~Sola, L.~Gonz\'alez:
  \emph{J. Phys. Chem. Lett.}, \textbf{3}, 3090 (2012).

\bibitem{Mai2013C}
S.~Mai, P.~Marquetand, M.~Richter, J.~Gonz\'alez-V\'azquez, L.~Gonz\'alez:
  \emph{ChemPhysChem}, \textbf{14}, 2920 (2013).

\bibitem{Richter2014PCCP}
M.~Richter, S.~Mai, P.~Marquetand, L.~Gonz\'{a}lez: \emph{Phys. Chem. Chem.
  Phys.}, \textbf{16}, 24423 (2014).

\bibitem{Tavernelli2011CP}
I.~Tavernelli, B.~F. Curchod, U.~Rothlisberger: \emph{Chem. Phys.},
  \textbf{391}, 101  (2011).

\bibitem{Freitag2014IC}
L.~Freitag, L.~Gonz\'alez: \emph{Inorg. Chem.}, \textbf{53}, 6415 (2014).

\bibitem{Richter2011JCTC}
M.~Richter, P.~Marquetand, J.~{Gonz\'alez-V\'azquez}, I.~Sola, L.~Gonz\'alez:
  \emph{J. Chem. Theory Comput.}, \textbf{7}, 1253 (2011).

\bibitem{Mai2014SHARC}
S.~Mai, M.~Richter, M.~Ruckenbauer, M.~Oppel, P.~Marquetand, L.~Gonz\'alez:
  SHARC: Surface Hopping Including Arbitrary Couplings -- Program Package for
  Non-Adiabatic Dynamics. sharc-md.org (2014).

\bibitem{Meyer2009}
H.-D. Meyer, F.~Gatti, G.~A. Worth: \emph{Multidimensional Quantum Dynamics}.
  Wiley-VCH Verlag GmbH \& Co. KGaA (2009).

\bibitem{Tannor2006}
D.~J. Tannor: \emph{Introduction to Quantum Mechanics: A Time-Dependent
  Perspective}. University Science Books (2006).

\bibitem{Burghardt1999JCP}
I.~Burghardt, H.-D. Meyer, L.~S. Cederbaum: \emph{J. Chem. Phys.},
  \textbf{111}, 2927 (1999).

\bibitem{Wang2003JCP}
H.~Wang, M.~Thoss: \emph{J. Chem. Phys.}, \textbf{119}, 1289 (2003).

\bibitem{Lasorne2006CPL}
B.~Lasorne, M.~J. Bearpark, M.~A. Robb, G.~A. Worth: \emph{Chem. Phys. Lett.},
  \textbf{432}, 604 (2006).

\bibitem{Martinez1996JPC}
T.~J. Mart\'{\i}nez, M.~Ben-Nun, R.~D. Levine: \emph{J. Phys. Chem.},
  \textbf{100}, 7884 (1996).

\bibitem{Tully1971JCP}
J.~C. Tully, R.~K. Preston: \emph{J. Chem. Phys.}, \textbf{55}, 562 (1971).

\bibitem{Tully1990JCP}
J.~C. Tully: \emph{J. Chem. Phys.}, \textbf{93}, 1061 (1990).

\bibitem{Doltsinis2002JTCC}
N.~L. Doltsinis, D.~Marx: \emph{J. Theor. Comput. Chem.}, \textbf{1}, 319
  (2002).

\bibitem{Barbatti2011WCMS}
M.~Barbatti: \emph{WIREs Comput. Mol. Sci.}, \textbf{1}, 620 (2011).

\bibitem{Persico2014TCA}
M.~Persico, G.~Granucci: \emph{Theor. Chem. Acc.}, \textbf{133}, 1526 (2014).

\bibitem{Malhado2014FC}
J.~P. Malhado, M.~J. Bearpark, J.~T. Hynes: \emph{Front. Chem.}, \textbf{2}, 97
  (2014).

\bibitem{Koppel1978JCP}
H.~K\"oppel, W.~Domcke, L.~S. Cederbaum, W.~von Niessen: \emph{J. Chem. Phys.},
  \textbf{69}, 4252 (1978).

\bibitem{Koeppel2007}
H.~K\"oppel, W.~Domcke, L.~S. Cederbaum: \emph{Multimode Molecular Dynamics
  Beyond the Born-Oppenheimer Approximation}. John Wiley \& Sons, Inc., 59--246
  (2007).

\bibitem{Henry1971JCP}
B.~R. Henry, W.~Siebrand: \emph{J. Chem. Phys.}, \textbf{54}, 1072 (1971).

\bibitem{Granucci2007JCP}
G.~Granucci, M.~Persico: \emph{J. Chem. Phys.}, \textbf{126}, 134114 (2007).

\bibitem{Granucci2010JCP}
G.~Granucci, M.~Persico, A.~Zoccante: \emph{J. Chem. Phys.}, \textbf{133},
  134111 (2010).

\bibitem{Cantatore2014CTC}
V.~Cantatore, G.~Granucci, M.~Persico: \emph{Comput. Theor. Chem.},
  \textbf{1040?1041}, 126 , excited states: From isolated molecules to complex
  environments Excited states (2014).

\bibitem{Bajo2014JCP}
J.~J. Bajo, G.~Granucci, M.~Persico: \emph{J. Chem. Phys.}, \textbf{140},
  044113 (2014).

\bibitem{Zhu2004JCP}
C.~Zhu, S.~Nangia, A.~W. Jasper, D.~G. Truhlar: \emph{J. Chem. Phys.},
  \textbf{121}, 7658 (2004).

\bibitem{Zhu2005JCTC}
C.~Zhu, A.~W. Jasper, D.~G. Truhlar: \emph{J. Chem. Theory Comput.},
  \textbf{1}, 527 (2005).

\bibitem{Jasper2005JCP}
A.~W. Jasper, D.~G. Truhlar: \emph{J. Chem. Phys.}, \textbf{122}, 044101
  (2005).

\bibitem{Jasper2007JCP}
A.~W. Jasper, D.~G. Truhlar: \emph{J. Chem. Phys.}, \textbf{127}, 194306
  (2007).

\bibitem{Cheng2008JCP}
S.~C. Cheng, C.~Zhu, K.~K. Liang, S.~H. Lin, D.~G. Truhlar: \emph{J. Chem.
  Phys.}, \textbf{129}, 024112 (2008).

\bibitem{Prezhdo1997JCPa}
O.~V. Prezhdo, P.~J. Rossky: \emph{J. Chem. Phys.}, \textbf{107}, 5863 (1997).

\bibitem{Jaeger2012JCP}
H.~M. Jaeger, S.~Fischer, O.~V. Prezhdo: \emph{J. Chem. Phys.}, \textbf{137},
  22A545 (2012).

\bibitem{Subotnik2011JCP}
J.~E. Subotnik, N.~Shenvi: \emph{J. Chem. Phys.}, \textbf{134}, 024105 (2011).

\bibitem{Shenvi2011JCP}
N.~Shenvi, J.~E. Subotnik, W.~Yang: \emph{J. Chem. Phys.}, \textbf{135}, 024101
  (2011).

\bibitem{Subotnik2013JCP}
J.~E. Subotnik, W.~Ouyang, B.~R. Landry: \emph{J. Chem. Phys.}, \textbf{139},
  214107 (2013).

\bibitem{Jakubetz1999PCCP}
W.~Jakubetz, J.~N.~L.~Connor, P.~J.~Kuntz: \emph{Phys. Chem. Chem. Phys.},
  \textbf{1}, 1213 (1999).

\bibitem{Bedard-Hearn2005JCP}
M.~J. Bedard-Hearn, R.~E. Larsen, B.~J. Schwartz: \emph{J. Chem. Phys.},
  \textbf{123}, 234106 (2005).

\bibitem{Nelson2013JCP}
T.~Nelson, S.~Fernandez-Alberti, A.~E. Roitberg, S.~Tretiak: \emph{J. Chem.
  Phys.}, \textbf{138}, 224111 (2013).

\bibitem{Yonehara2012CR}
T.~Yonehara, K.~Hanasaki, K.~Takatsuka: \emph{Chem. Rev.}, \textbf{112}, 499
  (2012).

\bibitem{Makri1989JCP}
N.~Makri, W.~H. Miller: \emph{J. Chem. Phys.}, \textbf{91}, 4026 (1989).

\bibitem{Hammes-Schiffer1994JCP}
S.~{Hammes-Schiffer}, J.~C. Tully: \emph{J. Chem. Phys.}, \textbf{101}, 4657
  (1994).

\bibitem{Takatsuka1999PR}
K.~Takatsuka, H.~Ushiyama, A.~Inoue-Ushiyama: \emph{Phys. Rep.}, \textbf{322},
  347  (1999).

\bibitem{Xing2001JPCB}
J.~Xing, E.~A. Coronado, W.~H. Miller: \emph{J. Phys. Chem. B}, \textbf{105},
  6574 (2001).

\bibitem{Larregaray2002PCCP}
P.~Larregaray, L.~Bonnet, J.-C. Rayez: \emph{Phys. Chem. Chem. Phys.},
  \textbf{4}, 1571 (2002).

\bibitem{Zener1932PRSA}
C.~Zener: \emph{Proc. R. Soc. A}, \textbf{137}, 696 (1932).

\bibitem{Granucci2012JCP}
G.~Granucci, M.~Persico, G.~Spighi: \emph{J. Chem. Phys.}, \textbf{137}, 22A501
  (2012).

\bibitem{Curchod2013C}
B.~F.~E. Curchod, T.~J. Penfold, U.~Rothlisberger, I.~Tavernelli:
  \emph{Chimia}, \textbf{67}, 218 (2013).

\bibitem{FrancodeCarvalho2014JCP}
F.~Franco~de Carvalho, B.~F.~E. Curchod, T.~J. Penfold, I.~Tavernelli: \emph{J.
  Chem. Phys.}, \textbf{140}, 144103 (2014).

\bibitem{Gai1992JAP}
H.~Gai, G.~Voth: \emph{J. Appl. Phys.}, \textbf{71}, 1415 (1992).

\bibitem{Mitric2009PRA}
R.~Mitri\'c, J.~Petersen, V.~Bona\v{c}\i\'c-Kouteck\'y: \emph{Phys. Rev. A},
  \textbf{79}, 053416 (2009).

\bibitem{Thachuk1996JCP}
M.~Thachuk, M.~Y. Ivanov, D.~M. Wardlaw: \emph{J. Chem. Phys.}, \textbf{105},
  4094 (1996).

\bibitem{Jones2008JPCA}
G.~A. Jones, A.~Acocella, F.~Zerbetto: \emph{J. Phys. Chem. A}, \textbf{112},
  9650 (2008).

\bibitem{Marquetand2011FD}
P.~Marquetand, M.~Richter, J.~Gonz\'alez-V\'azquez, I.~Sola, L.~Gonz\'alez:
  \emph{Faraday Discuss.}, \textbf{153}, 261 (2011).

\bibitem{Bajo2011JPCA}
J.~J. Bajo, J.~Gonz\'alez-V\'azquez, I.~Sola, J.~Santamaria, M.~Richter,
  P.~Marquetand, L.~Gonz\'alez: \emph{J. Phys. Chem. A}, \textbf{116}, 2800
  (2011).

\bibitem{Verlet1967PR}
L.~Verlet: \emph{Phys. Rev.}, \textbf{159}, 98 (1967).

\bibitem{Hess1996CPL}
B.~A. He{\ss}, C.~M. Marian, U.~Wahlgren, O.~Gropen: \emph{Chem. Phys. Lett.},
  \textbf{251}, 365 (1996).

\bibitem{Dyall2007}
K.~G. Dyall, K.~F{\ae}gri: \emph{Introduction to Relativistic Quantum
  Chemistry}. Oxford University Press (2007).

\bibitem{Reiher2009}
M.~Reiher, A.~Wolf: \emph{Relativistic Quantum Chemistry}. Wiley VCH Verlag
  Weinheim (2009).

\bibitem{Ballhausen1972ARPC}
C.~J. Ballhausen, A.~E. Hansen: \emph{Annu. Rev. Phys. Chem.}, \textbf{23}, 15
  (1972).

\bibitem{Kendrick2000CPL}
B.~K. Kendrick, C.~A. Mead, D.~G. Truhlar: \emph{Chem. Phys. Lett.},
  \textbf{330}, 629  (2000).

\bibitem{Baer2000CPL}
M.~Baer: \emph{Chem. Phys. Lett.}, \textbf{330}, 633  (2000).

\bibitem{Granucci2011JCC}
G.~Granucci, M.~Persico: \emph{J. Comput. Chem.}, \textbf{32}, 2690 (2011).

\bibitem{Mai2014JCP_reindex}
S.~Mai, T.~M\"uller, P.~Marquetand, F.~Plasser, H.~Lischka, L.~Gonz\'alez:
  \emph{J. Chem. Phys.}, \textbf{141}, 074105 (2014).

\bibitem{Lischka2012}
H.~Lischka, R.~Shepard, I.~Shavitt, R.~M. Pitzer, M.~Dallos, T.~M\"uller, P.~G.
  Szalay, F.~B. Brown, R.~Ahlrichs, H.~J. B\"ohm, A.~Chang, D.~C. Comeau,
  R.~Gdanitz, H.~Dachsel, C.~Ehrhardt, M.~Ernzerhof, P.~H\"ochtl, S.~Irle,
  G.~Kedziora, T.~Kovar, V.~Parasuk, M.~J.~M. Pepper, P.~Scharf, H.~Schiffer,
  M.~Schindler, M.~Sch\"uler, M.~Seth, E.~A. Stahlberg, J.-G. Zhao,
  S.~Yabushita, Z.~Zhang, M.~Barbatti, S.~Matsika, M.~Schuurmann, D.~R.
  Yarkony, S.~R. Brozell, E.~V. Beck, , J.-P. Blaudeau, M.~Ruckenbauer,
  B.~Sellner, F.~Plasser, J.~J. Szymczak: COLUMBUS, an ab initio electronic
  structure program, release 7.0 (2012).

\bibitem{Werner2012}
H.-J. Werner, P.~J. Knowles, G.~Knizia, F.~R. Manby, M.~{Sch\"{u}tz}, et~al.:
  MOLPRO, version 2012.1, a package of ab initio programs. www.molpro.net
  (2012).

\bibitem{Aquilante2010JCC}
F.~Aquilante, L.~De~Vico, N.~Ferr\'e, G.~Ghigo, P.-{\AA}. Malmqvist,
  P.~Neogr\'ady, T.~B. Pedersen, M.~Pito\v{n}\'ak, M.~Reiher, B.~O. Roos,
  L.~Serrano-Andr\'es, M.~Urban, V.~Veryazov, R.~Lindh: \emph{J. Comput.
  Chem.}, \textbf{31}, 224 (2010).

\bibitem{Granucci2001JCP}
G.~Granucci, M.~Persico, A.~Toniolo: \emph{J. Chem. Phys.}, \textbf{114}, 10608
  (2001).

\bibitem{Plasser2012JCP}
F.~Plasser, G.~Granucci, J.~Pittner, M.~Barbatti, M.~Persico, H.~Lischka:
  \emph{J. Chem. Phys.}, \textbf{137}, 22A514 (2012).

\bibitem{El-Sayed1963JCP}
M.~A. {El-Sayed}: \emph{J. Chem. Phys.}, \textbf{38}, 2834 (1963).

\bibitem{Albrecht1963JCP}
A.~C. Albrecht: \emph{J. Chem. Phys.}, \textbf{38}, 354 (1963).

\bibitem{Fabiano2008CP}
E.~Fabiano, T.~Keal, W.~Thiel: \emph{Chem. Phys.}, \textbf{349}, 334 , electron
  Correlation and Molecular Dynamics for Excited States and Photochemistry
  (2008).

\bibitem{Bearpark2007JPPA}
M.~J. Bearpark, F.~Ogliaro, T.~Vreven, M.~Boggio-Pasqua, M.~J. Frisch, S.~M.
  Larkin, M.~Morrison, M.~A. Robb: \emph{J. Photoch. Photobio. A},
  \textbf{190}, 207  (2007).

\bibitem{Olsen2011IJQC}
J.~Olsen: \emph{Int. J. Quant. Chem.}, \textbf{111}, 3267 (2011).

\bibitem{Martinez-Fernandez2014JCTC}
L.~Mart\'{\i}nez-Fern\'andez, J.~Gonz\'alez-V\'azquez, L.~Gonz\'alez,
  I.~Corral: \emph{J. Chem. Theory Comput.}, accepted (2014).

\bibitem{Corrales2014PCCP}
M.~E. Corrales, V.~Loriot, G.~Balerdi, J.~Gonz\'{a}lez-V\'{a}zquez,
  R.~de~Nalda, L.~Ba{\~n}ares, A.~H. Zewail: \emph{Phys. Chem. Chem. Phys.},
  \textbf{16}, 8812 (2014).

\bibitem{Pulay2011IJQC}
P.~Pulay: \emph{Int. J. Quant. Chem.}, \textbf{111}, 3273 (2011).

\bibitem{Mori2009CPL}
T.~Mori, S.~Kato: \emph{Chem. Phys. Lett.}, \textbf{476}, 97  (2009).

\bibitem{Kuhlman2012FD}
T.~S. Kuhlman, W.~J. Glover, T.~Mori, K.~B. Moller, T.~J. Mart\'inez:
  \emph{Faraday Discuss.}, \textbf{157}, 193 (2012).

\bibitem{Mori2012JPCA}
T.~Mori, W.~J. Glover, M.~S. Schuurman, T.~J. Mart\'{\i}nez: \emph{J. Phys.
  Chem. A}, \textbf{116}, 2808 (2012).

\bibitem{Nakayama2013JCP}
A.~Nakayama, G.~Arai, S.~Yamazaki, T.~Taketsugu: \emph{J. Chem. Phys.},
  \textbf{139}, 214304 (2013).

\bibitem{Marti2010ZPC}
K.~H. Marti, M.~Reiher: \emph{Z. Phys. Chem.}, \textbf{224}, 583 (2010).

\bibitem{Thiel2014WCMS}
W.~Thiel: \emph{WIREs Comput. Mol. Sci.}, \textbf{4}, 145 (2014).

\bibitem{Persico2003JMS-T}
M.~Persico, G.~Granucci, S.~Inglese, T.~Laino, A.~Toniolo: \emph{J. Mol. Struc.
  - {THEOCHEM}}, \textbf{621}, 119  (2003).

\bibitem{Senn2009ACIE}
H.~M. Senn, W.~Thiel: \emph{Angew. Chem. Int. Ed.}, \textbf{48}, 1198 (2009).

\bibitem{Hayashi2009BJ}
S.~Hayashi, E.~Tajkhorshid, K.~Schulten: \emph{Biophys. J.}, \textbf{96}, 403
  (2009).

\bibitem{Fingerhut2012JCP}
B.~P. Fingerhut, S.~Oesterling, K.~Haiser, K.~Heil, A.~Glas, W.~J. Schreier,
  W.~Zinth, T.~Carell, R.~de~Vivie-Riedle: \emph{J. Chem. Phys.}, \textbf{136},
  204307 (2012).

\bibitem{Pollard1992ARPC}
W.~T. Pollard, R.~A. Mathies: \emph{Annu. Rev. Phys. Chem.}, \textbf{43}, 497
  (1992).

\bibitem{Stolow2003ARPC}
A.~Stolow: \emph{Annu. Rev. Phys. Chem.}, \textbf{54}, 89 (2003).

\bibitem{Lowdin1950JCP}
P.~L\"owdin: \emph{J. Chem. Phys.}, \textbf{18}, 365 (1950).

\bibitem{Carlson1957PR}
B.~Carlson, J.~Keller: \emph{Phys. Rev.}, \textbf{105}, 102 (1957).

\end{thebibliography}

\end{document}